\documentclass[journal]{IEEEtran}

\usepackage{cite}
\usepackage{amsmath,amssymb,amsfonts}
\usepackage{algorithmic}
\usepackage{graphicx}
\usepackage{xcolor}
\usepackage{textcomp}
\usepackage{setspace}
\usepackage{pifont}
\usepackage{enumitem}
\usepackage{array}
\usepackage{longtable}
\usepackage{booktabs}

\setlist{nosep, topsep=0pt, partopsep=0pt, parsep=0pt, itemsep=1pt, leftmargin=*}

\def\BibTeX{{\rm B\kern-.05em{\sc i\kern-.025em b}\kern-.08em
T\kern-.1667em\lower.7ex\hbox{E}\kern-.125emX}}

\newcommand{\cmark}{\ding{51}}
\newcommand{\xmark}{\ding{55}}

\begin{document}

\title{The Rise of the Software-Defined Vehicle: Architectures, Enabling Technologies, and Future Opportunities}

\author{
Eirini Liotou, Dimitra Tzelalidou, and Gerasimos Christodoulou%
\thanks{The authors are with the Department of Informatics and Telematics,
Harokopio University of Athens, Greece.}%
\thanks{Corresponding author: Eirini Liotou (e-mail: eliotou@hua.gr).}
}

\maketitle
\begin{abstract}
The transition toward Software-Defined Vehicles (SDVs) represents a major paradigm shift in vehicle design, transforming traditional hardware-centric systems into software-centric platforms capable of dynamic adaptation and continuous functional evolution. SDVs enable advanced capabilities such as Over-the-Air (OTA) updates, intelligent automation, and connected services driven by AI. This paper presents a comprehensive survey of the architectures, enabling technologies, and operational frameworks that define modern SDVs. It examines the evolution of vehicle architectures from distributed electronic control unit (ECU) systems to domain-based, zonal, and centralized computing platforms. Key enabling technologies are reviewed, including service-oriented software architectures, middleware, automation pipelines, artificial intelligence mechanisms, and cloud-based infrastructures. A structured taxonomy is introduced to organize SDV technologies into functional hardware, E/E architectures, software frameworks, automation mechanisms, and distributed infrastructure domains. The study also investigates the Software-Defined Internet of Vehicles (SDIoV) paradigm, integrating Software-Defined Networking (SDN) with edge and fog computing to support scalable vehicular communication and data processing. Furthermore, key technical challenges related to cybersecurity, interoperability, data management, and system scalability are discussed, along with emerging research directions and future development trends.
\end{abstract}


\setlength{\abovecaptionskip}{5pt}
\setlength{\belowcaptionskip}{4pt}
\setlength{\floatsep}{8pt plus 2pt minus 2pt}
\setlength{\textfloatsep}{10pt plus 2pt minus 4pt}
\setlength{\intextsep}{8pt plus 2pt minus 2pt}
\setlength{\dblfloatsep}{8pt plus 2pt minus 2pt}
\setlength{\dbltextfloatsep}{10pt plus 2pt minus 4pt}

\setlength{\parskip}{0pt}
\setlength{\topsep}{0pt}
\setlength{\partopsep}{0pt}
\setlength{\itemsep}{0pt}
\setlength{\parsep}{0pt}

\section{Introduction}

\IEEEPARstart{T}{ransportation} has long been a fundamental component of modern society, enabling mobility, economic activity, and social connectivity. In the era of Hardware-Defined Vehicles (HDVs), vehicle functionality was primarily achieved through continuous improvements in mechanical design and the incremental addition of dedicated hardware components \cite{b1}. However, this traditional paradigm is undergoing a profound transformation. Modern vehicles are no longer viewed solely as mechanical systems that transport passengers from one location to another, but rather as intelligent platforms designed to enhance safety, comfort, connectivity, and user experience \cite{b2}.

In this context, the concept of Software-Defined Vehicles (SDVs) has emerged as a revolutionary approach in the automotive industry. SDVs transform vehicles from hardware-centric products into intelligent systems driven by advanced software architectures and programmable functionalities \cite{b8}. Core technologies supporting SDVs include automation, Over-The-Air (OTA) updates, infotainment systems, and advanced connectivity solutions, all of which contribute to enabling flexible, scalable, and continuously evolving vehicle capabilities \cite{b1,b29}.

Among these technologies, OTA updates allow manufacturers to remotely deploy software upgrades and security patches without physical access to the vehicle, reducing maintenance costs and enabling continuous feature evolution \cite{b55}. Infotainment systems integrate intelligent user interfaces and cloud-based services to provide navigation, entertainment, and communication, enhancing the in-vehicle experience \cite{b29}. Automation and ADAS technologies serve as intermediate steps toward fully autonomous operation, improving safety and driving efficiency \cite{Kukkala2018,b67,b66}. Connectivity via V2V communication enables real-time exchange of location, speed, and trajectory data, supporting coordinated decision-making and cooperative vehicular ecosystems \cite{Abdelkader2021,b104}.

As vehicular connectivity evolves, the SDIoV paradigm integrates SDN principles into vehicular communication infrastructures, enabling centralized network management, dynamic resource allocation, and improved Quality of Service (QoS) \cite{b94,b97,Anton2026}. SDV integration into the IoT ecosystem has increased data volumes and introduced challenges in processing, storage, and security \cite{b104,Haq25}. Architecturally, SDVs function as cyber-physical nodes within ITS, supporting V2X communication, cooperative driving, real-time traffic optimization, and continuous OTA-driven software evolution.

Motivated by these developments, this paper presents a comprehensive survey of SDVs and the SDIoV paradigm, clarifying their architectures, enabling technologies, challenges, and future directions.

The remainder is organized as follows: Section~II covers SDV architecture and enabling technologies; Section~III presents use cases and application domains; Section~IV introduces the SDIoV paradigm; Section~V discusses technical challenges; Section~VI concludes with future directions.

\subsection{The emergence of the SDV}
Vehicles have evolved from mechanically dominated machines into highly computerized platforms integrating hundreds of ECUs, diverse sensor technologies, and high-performance computing. Vehicle capabilities are now increasingly defined by software rather than hardware, reinforcing the transition toward the SDV paradigm \cite{b1,b2,b8,b55}. The term ``Software-Defined Vehicle'' (SDV) was introduced to describe this transition, which began with the gradual integration of ECUs and computing systems and has accelerated with the adoption of modern communication and software technologies.

\subsection{Related Work and Research Gap}
The literature on SDVs is growing, but it remains fragmented across publication venues and adjacent research communities. The limited number of journal-level SDV overviews published so far mainly introduces the paradigm, discusses the transition from distributed ECU-based designs toward centralized or zonal architectures, and outlines broad technical and organizational challenges rather than developing a unified analytical treatment of the full SDV ecosystem \cite{Stu25,Jol26}. In particular, \cite{Stu25} provides one of the clearest general overviews of SDV evolution and current challenges, while \cite{Jol26} discusses the engineering implications of SDVs for architecture design, software processes, update management, and compliance.

An explicit effort to structure the field can be found in recent review-oriented conference papers. The systematic review in \cite{Ott25} contributes to conceptual clarification and identifies future research avenues, while \cite{Mat25} attempts to bridge industry and academic perspectives by synthesizing competing definitions of SDVs. However, these works are primarily concerned with conceptualization and definitional alignment rather than with an integrated technical synthesis across in-vehicle computing, software platforms, communication systems, and distributed infrastructure. In parallel, security-focused surveys \cite{bN3,Sgh25} contextualise SDV attack surfaces, cybersecurity challenges, and privacy risks; \cite{bN3} additionally contributes an industry questionnaire across the automotive supply chain, and hybrid SDN--edge--cloud architectures have been proposed for QoS-aware vehicular networking in 6G environments \cite{Christ2025}. 

A mature body of adjacent literature surveys SDVNs, vehicular cloud, and SDN-enabled edge/fog computing \cite{Isl20,Mek21,Cardona2020,Nke23,Haq25,Lv21}, but focuses on network control rather than the vehicle as a software-defined cyber-physical platform integrating E/E architecture, OTA lifecycle management, AI services, and distributed orchestration.

Despite this progress, the reviewed literature remains fragmented across architecture, enabling software technologies, and cloud-edge / SDIoV perspectives, with no single work in our review offering a fully integrated analytical treatment spanning all of these dimensions \cite{Liu22,Pra23,Dam23}.

To address this gap, this paper presents a comprehensive and integrated survey spanning vehicle architecture, enabling software technologies, networking and cloud-edge infrastructures, and open system-level challenges. Table~\ref{tab:survey_comparison} positions this work against representative prior studies.

\begin{table*}[t]
\centering
\caption{Comparison of representative works. Columns correspond directly to contributions C1--C7.
\cmark\ = explicitly covered, $\sim$ = partially covered, \xmark\ = not a main focus.
\textsuperscript{\dag}\cite{Sgh25} is an unreviewed preprint; its coverage assessment reflects the version available at time of submission.}
\label{tab:survey_comparison}
\renewcommand{\arraystretch}{1.05}
\setlength{\tabcolsep}{2pt}
\scriptsize
\resizebox{\textwidth}{!}{%
\begin{tabular}{|l|c|l|l|c|c|c|c|c|c|c|c|}
\hline
\textbf{Ref.} &
\textbf{Year} &
\textbf{Venue} &
\textbf{Survey / review} &
\textbf{\begin{tabular}[c]{@{}c@{}}C1:\\ SDV Taxonomy\end{tabular}} &
\textbf{\begin{tabular}[c]{@{}c@{}}C2:\\ E/E Architecture\end{tabular}} &
\textbf{\begin{tabular}[c]{@{}c@{}}C3:\\ SW Frameworks$+$OTA\end{tabular}} &
\textbf{\begin{tabular}[c]{@{}c@{}}C4:\\ Automation$+$ AI\end{tabular}} &
\textbf{\begin{tabular}[c]{@{}c@{}}C5:\\ SDIoV / SDN / Edge-Fog\end{tabular}} &
\textbf{\begin{tabular}[c]{@{}c@{}}C6:\\ Challenges\end{tabular}} &
\textbf{\begin{tabular}[c]{@{}c@{}}C7:\\ Future Directions\end{tabular}} &
\textbf{\begin{tabular}[c]{@{}c@{}}Unified\\ View\end{tabular}} \\
\hline
\cite{Stu25} & 2025 & Journal    & Review / comprehensive study  & \cmark   & \cmark  & $\sim$  & $\sim$  & \xmark  & \cmark  & \cmark  & \xmark \\
\hline
\cite{Jol26} & 2026 & Journal    & Engineering overview          & $\sim$   & \cmark  & \cmark  & \xmark  & \xmark  & \cmark  & $\sim$  & \xmark \\
\hline
\cite{Ott25} & 2025 & Conference & Systematic review             & \cmark   & $\sim$  & \xmark  & \xmark  & \xmark  & $\sim$  & \cmark  & \xmark \\
\hline
\cite{Mat25} & 2025 & Conference & Review-oriented paper         & \cmark   & $\sim$  & \xmark  & \xmark  & \xmark  & $\sim$  & $\sim$  & \xmark \\
\hline
\cite{bN3}   & 2026 & Journal    & Survey                        & \cmark   & $\sim$  & \cmark  & \xmark  & \xmark  & \cmark  & $\sim$  & \xmark \\
\hline
\cite{Sgh25}\textsuperscript{\dag} & 2025 & Preprint   & Survey / taxonomy             & \cmark   & $\sim$  & \cmark  & \xmark  & \xmark  & \cmark  & $\sim$  & \xmark \\
\hline
\cite{MoToBFT} & 2026 & Journal    & Algorithm / approach          & \xmark   & $\sim$  & \xmark  & \xmark  & \cmark  & \cmark  & $\sim$  & \xmark \\
\hline
\cite{Isl20} & 2021 & Journal    & Survey                        & \xmark   & \xmark  & \xmark  & \xmark  & \cmark  & \cmark  & $\sim$  & \xmark \\
\hline
\cite{Mek21} & 2022 & Journal    & Survey                        & \xmark   & \xmark  & \xmark  & \xmark  & \cmark  & \cmark  & $\sim$  & \xmark \\
\hline
\cite{Nke23} & 2023 & Journal    & Survey                        & \xmark   & \xmark  & \xmark  & \xmark  & \cmark  & \cmark  & $\sim$  & \xmark \\
\hline
\cite{Haq25} & 2025 & Journal    & Survey                        & \xmark   & \xmark  & \xmark  & $\sim$  & \cmark  & \cmark  & \cmark  & \xmark \\
\hline
\textbf{This work} & \textbf{2026} & \textbf{--} & \textbf{Comprehensive survey} &
\textbf{\cmark} & \textbf{\cmark} & \textbf{\cmark} & \textbf{\cmark} &
\textbf{\cmark} & \textbf{\cmark} & \textbf{\cmark} & \textbf{\cmark} \\
\hline
\end{tabular}%
}
\end{table*}

\subsection{Survey contributions}
Motivated by the fragmentation of existing studies across
individual technical domains, this survey aims to provide
a structured overview of SDV technologies by connecting
architectural design, enabling software components, automation
processes, and distributed infrastructure systems within a
unified analytical context.

Unlike many prior works that focus on isolated aspects of
software-defined mobility (e.g., networking, security, or
architectural evolution), this paper emphasizes the relationships
among multiple components of the SDV ecosystem and presents
a structured description of their functional interactions.
The main contributions of this survey are summarized as follows,
and are explicitly mapped to the columns of Table~\ref{tab:survey_comparison}:

\begin{itemize}

\item[\textbf{C1}]
\textbf{Structured taxonomy and SDV concept:}
This paper consolidates SDV definitions from the literature
and organizes SDV technologies into five domains:
functional hardware, E/E architectures, software frameworks,
automation mechanisms, and distributed infrastructure.

\item[\textbf{C2}]
\textbf{Description of SDV architectural evolution:}
The survey presents the transition from distributed
ECU-based systems to domain-based, zonal,
and centralized E/E architectures, highlighting the
motivations behind these developments.

\item[\textbf{C3}]
\textbf{Systematic overview of enabling software technologies:}
The paper discusses operating systems, middleware,
service-oriented architectures, and OTA update mechanisms
within a unified SDV architectural framework.

\item[\textbf{C4}]
\textbf{Structured automation and AI analysis:}
Automation functions are organized into a coherent
pipeline including detection, positioning, mapping,
motion planning, and vehicle control, and the AI techniques
underpinning each stage are reviewed.

\item[\textbf{C5}]
\textbf{Extension toward SDIoV infrastructures:}
The survey extends the discussion from in-vehicle
systems to SDIoV architectures, integrating SDN,
edge computing, and fog computing with vehicular platforms.

\item[\textbf{C6}]
\textbf{Analysis of key technical challenges:}
Major challenges including cybersecurity,
interoperability, scalability, data management,
and energy efficiency are identified and discussed.

\item[\textbf{C7}]
\textbf{Discussion of future research directions:}
The paper outlines emerging trends related to
digital twins, federated learning, proactive cybersecurity,
and AI-defined vehicles.

\end{itemize}

Table~\ref{tab:survey_comparison}
summarizes the relationship between the above contributions
and representative prior works.

\subsection{Taxonomy of SDV Technologies}
\label{subsec:taxonomy}

\begin{figure*}[t]
\centering
\includegraphics[width=0.8\textwidth]{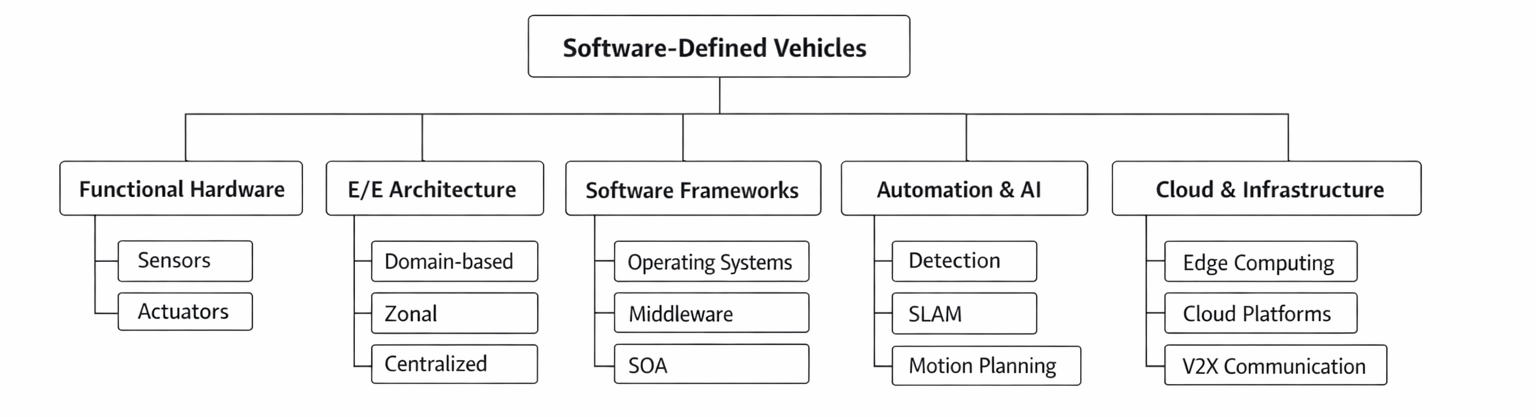}
\caption{Proposed taxonomy of SDV technologies, organizing the SDV ecosystem into functional hardware, E/E architectures, software frameworks, automation mechanisms, and distributed infrastructure domains.}
\label{fig:sdv_taxonomy}
\end{figure*}

To provide a structured overview of the SDV ecosystem, this survey organizes the core technologies and functional components into a unified taxonomy, as illustrated in Fig.~\ref{fig:sdv_taxonomy}. 

The proposed taxonomy groups SDV technologies into five primary domains: (i) Functional Hardware and Sensing Systems, (ii) Electrical/Electronic (E/E) Architectures, (iii) Software and Service-Oriented Frameworks, (iv) Automation and Artificial Intelligence Mechanisms, and (v) Cloud and Distributed Infrastructure Platforms. 

Each domain represents a critical layer of the SDV ecosystem, enabling modular development, scalability, and integration of advanced functionalities. This structured categorization facilitates systematic analysis of technologies, highlights their interdependencies, and supports the identification of open challenges and research directions across the SDV lifecycle.

\subsection{Definitions of the SDV}
Despite the growing prominence of the SDV in both academic and industrial discourse, the literature does not yet provide a single universally accepted definition. The term SDV is often confused with various concepts that have appeared from time to time for vehicles, such as automatic, electric, self-driving, connected and other variations. The term began to establish itself around the early 2020s and was created to describe the evolution of cars from simple mechanical devices to integrated computer systems with functions that are software-enabled. Existing studies describe SDV through a set of closely related perspectives that emphasize the increasing centrality of software in determining vehicle functionality, performance, upgradeability, and user experience \cite{Mat25,Jiang2024,Ott25}. This definitional diversity reflects the fact that SDV is not merely a new technical term, but a broader concept emerging from the transformation of the automotive industry toward software-centric development.

Across the literature, an SDV is consistently described as a vehicle in which software is the primary mechanism for implementing, modifying, and extending vehicle functions over its lifecycle \cite{Jiang2024,b1}. Enabling architectural characteristics—hardware--software decoupling, centralized or zonal computing, middleware platforms, pervasive connectivity, and OTA capability—are best understood as technological foundations rather than the definition itself \cite{Jiang2024, Mat25,Ott25}. Beyond the technical dimension, SDV also represents an industrial transformation, enabling post-sale feature delivery, user adaptation, and lifecycle value creation \cite{Mat25}.

After a thorough analysis of the various definitions proposed for SDV, the following comprehensive definition is proposed: ``SDV is an in-vehicle platform that enables the abstraction and software-based management of vehicle hardware functions, creating an extensible architecture with centralized controls. In addition, all vehicle software components must support OTA updates and have high standards of security and reliability.'' This definition captures the core philosophy of SDV design by combining the key characteristics identified above, namely software centricity, hardware decoupling, OTA capability, scalability, and security.

\subsection{Characteristics}
Based on the previous definitions, in order for a vehicle to qualify as an SDV, it must exhibit two foundational properties \cite{Jiang2024,b8}:

\begin{itemize}
    \item \textbf{Software-centric approach:} Software has the central role in the operation and capabilities of the vehicle, as all physical components of the vehicle (such as the engine, sensors, and processors) must be managed and controlled by software \cite{Jiang2024}.
    \item \textbf{Central control:} Vehicle control is assigned to a high-performance central computing platform that manages large volumes of data, coordinates subsystems, and integrates new functions without degrading system performance \cite{Jiang2024,Mat25}.
\end{itemize}

These two properties are the distinguishing markers of the SDV paradigm. The remaining architectural enablers—OTA update capability, hardware--software decoupling, scalability through cloud connectivity, and compliance with functional safety and cybersecurity standards (ISO~26262, ISO/SAE~21434)—follow as necessary technical consequences of software-centric, centrally controlled design and are examined in detail in Section~\ref{sec:arch} \cite{b20,bN2,bN4}.

\section{Architecture and Enabling Technologies of SDVs}
\label{sec:arch}
Rapid technological developments have forced the automotive industry to redefine the vehicle architecture, moving from traditional structures to modern, flexible, and connected platforms. This section focuses on the technical elements that make up the reshaping of vehicle architecture, analyzing functional hardware, electrical and electronic (E/E) system architecture, software architecture, service-oriented communication (SOC), applications, and cloud service platforms.

\begin{figure}[!t] 
\centering 
\includegraphics[width=0.5\textwidth]{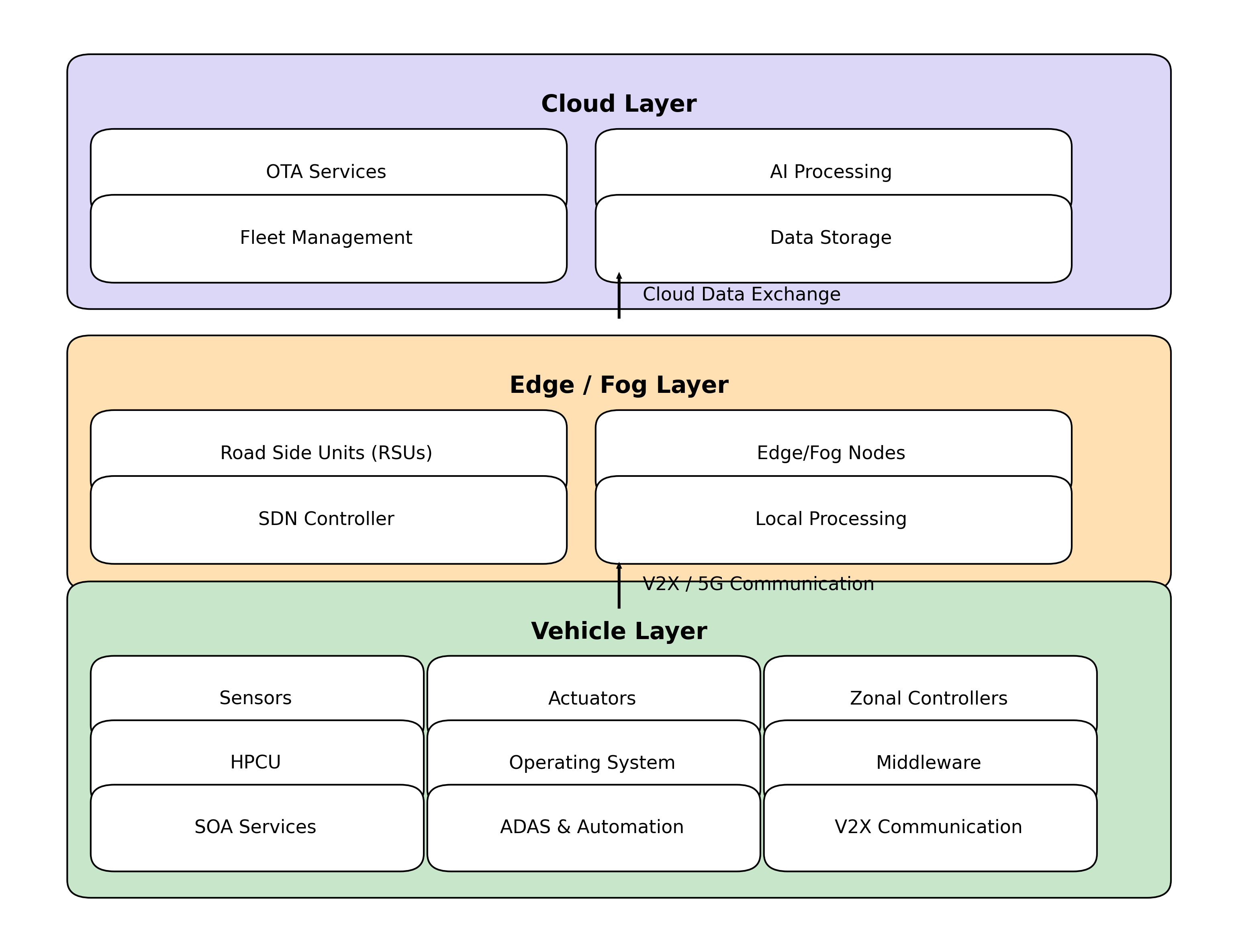}
\caption{Layered reference architecture of SDVs, illustrating interactions among in-vehicle components, edge/fog infrastructure, and cloud services supporting V2X communication and OTA-enabled functionality.} 
\label{fig:sdv_architecture_new} 
\end{figure}

Figure~\ref{fig:sdv_architecture_new} illustrates a layered reference architecture of SDVs, highlighting the interaction between in-vehicle hardware and software components, edge/fog infrastructure, SDN-enabled networking, cloud services, and V2X communication links.

\subsection{Technical elements of the architecture}
The architecture of cars, which is still used in some vehicles, is the traditional human-driven architecture. In this model, cars have independent systems, minimal connection to the external environment, and reduced functionality in terms of modern automation requirements, as they do not have the necessary software infrastructure. In the traditional architecture, the software is inextricably linked to the hardware and, in particular, to specialized ECUs. Such a vehicle increased its capabilities by adding more ECUs. In the most advanced cars, there may be over 100 ECUs to cover all functionality \cite{b53}. However, this architecture faces some problems, one of which is the complexity of the wiring, which makes it difficult to modify or upgrade the vehicle after purchase \cite{b53,Jol26}.

The redesign of automotive architecture is clearly necessary in order to meet these new requirements and enable scalability and management of large volumes of data without restrictions. This need inspired the creation of the SDV architecture. A well designed concept that harmoniously integrates hardware, software, and external systems in an efficient manner \cite{b20,Jiang2024}. In order to fully understand how SDVs function, the technical details of their architecture will be analyzed in detail below. This analysis will showcase how the coherent architecture of SDVs contributes to the creation of a flexible, scalable, safe, reliable, efficient, and user-friendly system.

Table~\ref{tab:sdv_arch_evolution} summarizes the transition from traditional ECU-based designs to centralized SDV architectures.

\begin{table*}[!t]
\caption{Evolution of vehicle E/E architectures toward SDVs}
\label{tab:sdv_arch_evolution}
\centering

\footnotesize
\renewcommand{\arraystretch}{0.82}
\setlength{\tabcolsep}{4pt}
\setlength{\extrarowheight}{0pt}

\begin{tabular*}{\textwidth}{@{\extracolsep{\fill}}|p{3.8cm}|p{4.3cm}|p{4.3cm}|p{4.3cm}|}
\hline
\textbf{Architecture} & \textbf{Main Characteristics} & \textbf{Advantages} & \textbf{Limitations} \\
\hline

Traditional ECU-based architecture
&
Large number of dedicated ECUs, tight hardware--software coupling, function-specific control units, distributed wiring harnesses
&
Mature design approach, modular function deployment, proven use in conventional vehicles
&
High wiring complexity, limited flexibility, difficult upgrades, integration overhead, poor scalability
\\
\hline

Domain-based architecture
&
ECUs grouped by functional domains such as ADAS, infotainment, and powertrain, partial consolidation of control logic
&
Reduced ECU proliferation, improved manageability, better functional grouping
&
Still complex to integrate across domains, limited global optimization, transition complexity
\\
\hline

Zonal architecture
&
Controllers organized according to physical zones of the vehicle, local aggregation of sensors and actuators, high-speed backbone communication
&
Reduced wiring length, lower weight, simplified power distribution, improved scalability, easier integration of new functions
&
Requires careful coordination between zones, strong communication backbone, higher software orchestration complexity
\\
\hline

Centralized SDV architecture
&
Central computing platform, software-driven control, service-oriented design, hardware--software decoupling, support for OTA updates and cloud connectivity
&
High flexibility, continuous software evolution, simplified updates, improved resource sharing, strong support for advanced automation
&
High computing demands, strict real-time requirements, cybersecurity concerns, increased software complexity
\\
\hline

\end{tabular*}
\end{table*}

\subsubsection{Functional hardware}

The term functional hardware describes all the hardware on the vehicle, such as advanced sensors (high-resolution cameras, high-performance radar, LiDAR, etc.) and actuators. Sensors collect data in real time; actuators convert electrical information into mechanical action to enable features such as collision avoidance, lane changing, and automatic positioning \cite{b8,b29,b33}. Table~\ref{tab:sensor_comparison} summarizes the most commonly used sensing technologies. Cameras support visual perception and ADAS functions; radar provides reliable distance and velocity estimation under adverse weather \cite{b33}. LiDAR enables high-precision 3D spatial mapping; hybrid sensor fusion outperforms any single modality, though cost remains a limitation. Ultrasonic sensors cover short-range proximity detection, and GNSS/IMU systems support localization and motion estimation \cite{Stu25}. Temperature, pressure, and torque sensors support vehicle-state monitoring \cite{b53,Stu25,b1}.

\begin{table*}[t]
\caption{Comparison of sensing technologies used in SDVs}
\label{tab:sensor_comparison}
\centering

\footnotesize
\renewcommand{\arraystretch}{0.76}
\setlength{\tabcolsep}{3pt}
\setlength{\extrarowheight}{-0.5pt}

\begin{tabular*}{\textwidth}{@{\extracolsep{\fill}}|p{3.5cm}|p{4.0cm}|p{4.5cm}|p{4.5cm}|}
\hline
\textbf{Sensor} & \textbf{Main Role} & \textbf{Advantages} & \textbf{Limitations} \\
\hline

Camera & Visual perception and object recognition & High-resolution imaging, supports ADAS and driver monitoring & Sensitive to lighting and weather conditions \\
\hline
Radar & Distance and velocity estimation & Reliable in adverse weather and long-range detection & Lower spatial resolution compared to cameras and LiDAR \\
\hline
LiDAR & 3D environment sensing & High precision and accurate spatial mapping & High cost and integration complexity \\
\hline
Ultrasonic & Short-range object detection & Low cost and effective at close distances & Limited sensing range \\
\hline
GNSS / IMU & Localization and motion estimation & Supports positioning and navigation tasks & Reduced accuracy in tunnels or dense urban areas \\
\hline
Temperature / Pressure / Torque sensors & Vehicle-state monitoring & Support system regulation and diagnostics & Limited environmental perception capability \\
\hline
\end{tabular*}
\end{table*}

Actuators translate software commands into mechanical actions. Electric actuators adjust mirror positions and window operation; hydraulic actuators control tire steering direction; pneumatic actuators manage brake pressure and braking distance; and piezoelectric actuators regulate engine fuel injection for improved combustion efficiency \cite{b37}.

\subsubsection{Electrical/Electronic Systems Architecture (E/E Architecture)}

Modern vehicle electrical/electronic (E/E) architectures have evolved significantly in response to increasing computational demands and system complexity. Early vehicle systems relied on distributed ECUs, where each function was managed independently. As functionality increased, domain-based architectures emerged to group related functions under centralized domain controllers. More recently, zonal and centralized architectures have been introduced to reduce wiring complexity, improve scalability, and enable software-driven vehicle functionality. This architectural evolution forms the technological foundation of SDVs; the main characteristics, advantages, and limitations of these architecture types are summarized in Table~\ref{tab:sdv_arch_evolution}.

As discussed above, the increasing functionality of cars requires greater computing power, which leads to an increase in the number of embedded ECUs. Therefore, such architectures create complexity in system organisation, as updating and compatibility between systems is not easily manageable \cite{b20}.

To address the limitations of fully distributed systems, domain-based architectures were introduced. In this approach, ECUs are grouped into functional domains such as powertrain, body control, infotainment, and ADAS. Each domain is managed by a domain controller responsible for coordinating the operation of ECUs within that functional area. This reduces communication overhead between independent ECUs and improves system organisation. However, communication between domains still requires complex network management, and scalability limitations remain when new functionalities are added.

For these reasons, scientists in the field have proposed a transition to more centralized architectures, leading to the development of zonal architecture for electrical/electronic (E/E) systems (Zonal Oriented Architecture). Zonal Architecture represents a progressive approach to the design and layout of electrical and electronic (E/E) systems in vehicles. Instead of the traditional structure with multiple independent control units decentralized throughout the vehicle, Zonal Architecture proposes grouping these and the functional materials (radar, lidar, camera) into areas or ``zones'' based on their physical location in the vehicle \cite{Jiang2024,Stu25,Jol26}.

\begin{figure}[!ht]
\centering
\includegraphics[width=0.48\textwidth]{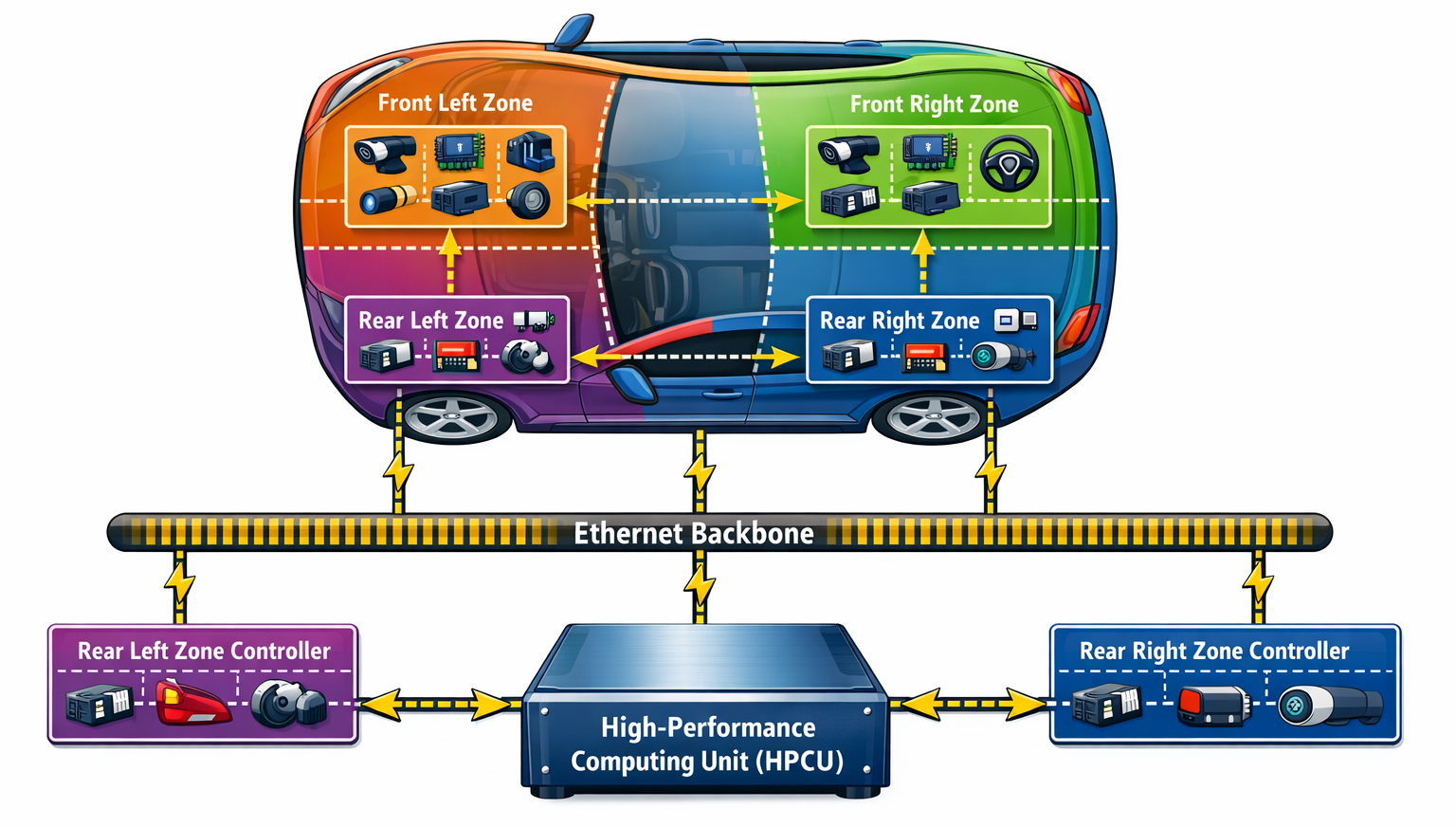}
\caption{Illustration of the zonal architecture principle, where vehicle components are grouped into physical zones managed by zone controllers connected to a central high-performance computing unit (HPCU) through a high-speed Ethernet backbone.}
\label{fig:zonal}
\end{figure}

The zonal architecture principle is illustrated in Fig.~\ref{fig:zonal}, where the vehicle is divided into multiple physical zones connected to a central computing platform.

As shown in Fig.~\ref{fig:zonal}, each zone is equipped with a zone controller that is responsible for managing the actuators, sensors, and ECUs located in a local sub-area of the vehicle (e.g., front right or left or rear of the vehicle). Each zone controller provides controlled power supply to the ECUs, sensors, and actuators in the corresponding zone. These controllers are connected to each other and to a high-performance computing unit (HPCU) via very fast Ethernet networks that enable data transmission at Gbps, which common bus communication protocols (such as Controller Area Network (CAN), Local Interconnect Network (LIN) and FlexRay) cannot offer. This structure ensures intelligent task distribution between all levels.

The HPCU, or high-performance computing platform, is located at the heart of the vehicle, performing critical tasks for its smooth operation. It handles the transfer of large amounts of data between the vehicle's internal and external networks, manages resources, facilitates intelligent decision-making, and monitors system safety and performance \cite{b53}.

Despite the coordination complexity, zonal architecture reduces ECU count, minimizes wiring weight and cost, simplifies communication networks, and eases integration of new technologies \cite{Jiang2024,b45,Stu25}.

\subsubsection{Software Architecture}

In the context of SDVs, software architecture is built upon multiple interacting layers that enable flexibility, modularity, and efficient system integration. In particular, three key elements emerge as the foundations of software architecture: the operating system, middleware, and the SOA (Service-Oriented Architecture) service layer. This approach offers a new, more distributed framework that allows for the efficient management of new data. The operating system provides the basis for running applications that are critical to vehicle operation and user experience, and manages and controls physical resources (such as sensors, etc.). The middleware acts as a link between the operating system and the applications. Its main role is to ensure proper interaction between systems, while providing flexibility to the vehicle. On the other hand, SOA presents a new approach to software based on modular services designed to work together and interact while acting independently (from each other, the platform, the language, etc.) \cite{b8,b52}.

\textbf{Operating System (OS)}: The concept of a vehicle operating system represents the entire software executed in kernel mode. It consists of two components: the Base Layer and the Software Factory \cite{b53}. The Base Layer is the basic software infrastructure executing main vehicle functions. The Software Factory is a dynamic environment that facilitates update automation and promotes the development of new applications and their integration into the rest of the system \cite{b53}. The OS is involved in all stages of vehicle development, from initial design to final market release.

The vehicle operating system now serves as a flexible platform that supports a wide range of functions in various areas within the vehicle. Unlike traditional operating systems, it extends its reach to cover sensor management for data collection in IoT applications, support for embedded systems, and real-time operating systems (RTOS), broadly in line with the architectural trends described in \cite{b53}. Across the wider embedded and IoT ecosystem, this functional reach is commonly organized into three illustrative categories:
\begin{itemize}
    \item \textbf{IoT/Sensor Node OS:} Lightweight systems such as TinyOS, Contiki, MantisOS, and LiteOS typically follow event-based programming models for efficiency and responsiveness in handling asynchronous events, while sharing common challenges such as optimizing energy consumption and implementing security mechanisms to safeguard sensitive data.

    \item \textbf{Embedded systems OS:} Platforms such as uCLinux, Raspbian, QNX Neutrino, Android Things, and WindRiver VxWorks perform specific functions within a larger electronic or mechanical system, typically prioritizing real-time response and fault tolerance given their connection to critical infrastructure.

    \item \textbf{RTOS (Real Time Operating System):} Systems such as FreeRTOS (open source, highly portable, low-power kernel), RIOT (open source, high-performance interrupt and memory management), and Nucleus RTOS (designed for high safety requirements with multicore SoC support) provide deterministic execution guarantees, task scheduling, priority management, and interrupt handling.
\end{itemize}
These examples are illustrative of categories widely discussed in the embedded-systems literature rather than an exhaustive or formally benchmarked list.

OEM cooperation is generally considered important for developing a common OS reference language that enables cross-manufacturer compatibility and a unified innovation ecosystem \cite{b53}.

\textbf{Service-Oriented Architecture (SOA)}: SOA encapsulates software components as interoperable, loosely coupled services that can independently produce or consume data \cite{b50}. Key advantages for automotive systems: modular flexibility (functions can be added or removed without full system rebuilds), personalization (drivers can activate/deactivate features on demand), simplified updates and error correction (OTA without workshop visits), reduced hardware-change costs, and service reusability across vehicle domains \cite{b50,b52,b47}.

\textbf{Middleware}: Middleware sits between hardware/OS and application software, abstracting hardware details, enabling virtualization, and ensuring interoperability across zone controllers \cite{b52,b50}. The dominant automotive middleware is \textbf{SOME/IP} (AUTOSAR), which provides publish/subscribe, request/response, and fire-and-forget patterns with selective data transfer, service discovery, and event notification over IP-based networks \cite{b50,b52}.

\textbf{Service-Oriented Communication (SOC)}: SOC enables dynamic Ethernet-based communication among \textit{service providers} (e.g., sensors publishing wheel speed or temperature), \textit{service consumers} (e.g., ADAS reading sensor data), a fault-tolerant \textit{middleware bus}, and \textit{control services} that monitor and coordinate all interactions \cite{b50,b52}. Table~\ref{tab:software_architecture_components} summarizes these main software architecture components and their roles.

\begin{table*}[t]
\caption{Main software architecture components in SDVs}
\label{tab:software_architecture_components}
\centering
\renewcommand{\arraystretch}{1.1}
\setlength{\tabcolsep}{4pt}
\footnotesize
\begin{tabular*}{\textwidth}{@{\extracolsep{\fill}}|p{3.5cm}|p{4.5cm}|p{4.5cm}|p{4.5cm}|}
\hline
\textbf{Component} & \textbf{Main Role} & \textbf{Key Characteristics} & \textbf{Representative Examples / Functions} \\
\hline
Operating System (OS) & Executes core vehicle software and manages hardware resources & Supports application execution, resource control, embedded functionality, and real-time response & TinyOS, Contiki, LiteOS, uCLinux, QNX, FreeRTOS, RIOT, Nucleus RTOS \\
\hline
Middleware & Acts as an intermediate layer between hardware, OS, and applications & Enables interoperability, abstraction, virtualization, communication, and flexible software integration & SOME/IP, service discovery, request/response, publish/subscribe \\
\hline
Service-Oriented Architecture (SOA) & Organizes software into modular and interoperable services & Loose coupling, modularity, reusability, flexibility, and independent service interaction & Distributed software services across vehicle domains \\
\hline
Service-Oriented Communication (SOC) & Supports dynamic communication among system elements over the network & Service providers, service consumers, middleware support, and control services & Ethernet-based communication, data exchange between sensors, ADAS, and control modules \\
\hline
\end{tabular*}
\end{table*}

\subsubsection{Applications}
For explanatory purposes, vehicle applications can be grouped into \textbf{functional} (safety-critical hardware control: FCW, AEB \cite{Jiang2024}) and \textbf{service} (entertainment, ergonomics, navigation: seat adjustment, music, weather \cite{b47}) categories, each running natively on the vehicle platform.

\subsubsection{Cloud Service Platform}
Sensor data flows from zone controllers through the HPCU to the TCU, which transmits it via 5G to cloud platforms for AI/ML-based processing and analysis \cite{b47,b50,Bordoloi2023}. Core cloud platform functions include:
\begin{itemize}
    \item \textbf{Data storage and processing}: Data storage is dynamically adapted to current needs, while powerful processing power is capable of handling the large volumes of data generated by vehicles. \cite{b47,b50}
    \item \textbf{Real-time updates}: OTA updates facilitate software updates and upgrades through techniques for finding and correcting errors or omissions, as well as adding new features remotely, keeping vehicles up to date and operational. \cite{b47}
    \item \textbf{Security}: The cloud platform protects vehicle data from threats and breaches. Encryption, authentication, and access control are critical factors it offers to maintain data integrity and confidentiality. \cite{bN2,bN4}
\end{itemize}
Major commercial cloud platforms support SDV deployments through device management with scalable compute and storage \cite{b47}, cloud connectivity with data integration and analytics services \cite{b47}, and managed machine-learning pipelines \cite{b50,Bordoloi2023}.

\subsection{Key characteristics}
At the core of SDVs are three fundamental features: OTA updates, automation, and Artificial Intelligence (AI). OTA updates enable real-time modifications to vehicle software, automation introduces advanced systems that manage various driving functions, and AI further enhances these capabilities, enabling intelligent decision-making, predictive maintenance, and personalized user experiences. Together, these features define the next generation of vehicles, offering unprecedented levels of performance, safety, and intelligence.

\subsubsection{Over-the-Air Updates}

OTA updates have emerged as a key feature in the evolution of SDVs. This approach allows vehicles to receive software upgrades and fixes wirelessly, just as smart devices update their apps and operating systems. By leveraging this dynamic update process, vehicles evolve throughout their lifecycle, improving performance and safety mechanisms in real time without the need for physical intervention \cite{b55}. OTA updates divide into two main categories: firmware over-the-air (FOTA) and software over-the-air (SOTA) updates. Table~\ref{tab:ota_comparison} compares FOTA and SOTA across update scope, system impact, testing, deployment complexity, and typical use cases.

\begin{table*}[t]
\caption{Comparison between Firmware Over-the-Air (FOTA) and Software Over-the-Air (SOTA) updates}
\label{tab:ota_comparison}
\centering
\renewcommand{\arraystretch}{1.1}
\setlength{\tabcolsep}{4pt}
\footnotesize
\begin{tabular*}{\textwidth}{@{\extracolsep{\fill}}|p{4.5cm}|p{6.0cm}|p{6.0cm}|}
\hline
\textbf{Characteristic} & \textbf{FOTA} & \textbf{SOTA} \\
\hline
Update Scope & Core firmware and operating system components & Application-level software components \\
\hline
System Impact & Affects critical vehicle functions such as ECUs and control units & Affects non-critical functions such as infotainment and navigation \\
\hline
Testing Requirements & Requires extensive validation and safety checks & Requires less rigorous testing compared to FOTA \\
\hline
Deployment Complexity & Higher complexity due to system dependencies & Lower complexity and faster deployment cycles \\
\hline
Typical Use Cases & Safety patches, ECU updates, system firmware upgrades & Feature additions, UI updates, infotainment improvements \\
\hline
\end{tabular*}
\end{table*}

\begin{itemize}
\item \textbf{Firmware Over-the-Air (FOTA) Updates}: FOTA updates include modifications to the vehicle's existing operating system and core system software. These updates are critical because they affect multiple applications managing integral parts of the vehicle's central operation such as ECUs, braking, and steering. FOTA allows manufacturers to push updates to specific units without disrupting the rest, minimizing downtime. FOTA usually includes a rollback mechanism if the new firmware introduces problems \cite{b55}.
\item \textbf{Software Over-the-Air (SOTA) Updates}: SOTA updates focus on the application level, allowing software upgrades without changing the core OS. These updates improve entertainment systems, navigation software, and user interfaces. They are generally less complex than FOTA, require less rigorous testing, and are released more quickly \cite{b55}.
\end{itemize}

The OTA update process comprises two phases. \textbf{Phase 1 (software module management)} covers: (i)~binary compilation using hardware-image, dynamic-linking, or pre-linking methods; (ii)~compatibility analysis to verify module--hardware and inter-module coexistence; and (iii)~behavior verification via sandboxes, simulators, or digital twins \cite{b55,bN2,bN4}. \textbf{Phase 2 (secure deployment)} covers: (i)~HMAC-based authentication and integrity checking during transmission; (ii)~broadcast/multicast dissemination with packet-fragmentation handling; and (iii)~installation and coordinated activation with automatic rollback on failure \cite{b55,bN2,bN4}. Key challenges include secure authentication, bandwidth optimization during large-scale rollouts, multi-version coexistence, and the stringent safety validation overhead of FOTA updates.

\subsubsection{Automation Systems}

Advanced automation systems reduce driver intervention through a structured pipeline covering detection, positioning, mapping, motion planning, and vehicle control. Table~\ref{tab:automation_pipeline} summarizes the core functional stages of this pipeline.

\begin{table*}[t]
\caption{Core functional stages of automation systems in SDVs}
\label{tab:automation_pipeline}
\centering
\renewcommand{\arraystretch}{1.1}
\setlength{\tabcolsep}{4pt}
\footnotesize
\begin{tabular*}{\textwidth}{@{\extracolsep{\fill}}|p{3.5cm}|p{4.5cm}|p{4.5cm}|p{4.0cm}|}
\hline
\textbf{Stage} & \textbf{Main Function} & \textbf{Key Technologies} & \textbf{Output} \\
\hline
Detection & Collect environmental data and identify objects & Cameras, Radar, LiDAR, Ultrasonic sensors & Object detection and scene understanding \\
\hline
Positioning & Determine vehicle location and orientation & GNSS, IMU, RTK, dead reckoning & Vehicle coordinates and trajectory reference \\
\hline
Mapping / SLAM & Build and update environment representation & HD maps, SLAM algorithms, GIS systems & Digital representation of environment \\
\hline
Motion Planning & Generate safe vehicle trajectory & Global and local planning algorithms & Collision-free route \\
\hline
Vehicle Control & Execute planned motion & Drive-by-wire, ACC, ESP systems & Controlled speed and steering \\
\hline
\end{tabular*}
\end{table*}

\begin{enumerate}
    \item Detection: This is the first step in vehicle automation, involving the use of various sensors to collect data about the vehicle's environment. The data is collected from a wide range of sources, such as visual information, object distance and speed values, and high-resolution 3D representations of space. The more relevant the features are, the closer we will be to understanding the current traffic situation and the overall picture of the road network \cite{b66,b65}.
    \item Positioning: Positioning in autonomous vehicles is a key aspect of navigation, determining the position and orientation of the vehicle relative to a global coordinate system, such as the World Geodetic System (WGS 84). While global navigation satellite systems (GNSS), such as GPS, provide absolute coordinates, the road network is a case where, due to conditions, weak frequency transmissions are often created, leading to a degradation in accuracy. To mitigate these errors, advanced techniques are used that integrate data from GNSS receivers with data from high-precision sensors, such as inertial navigation units (IMU), real-time kinematic positioning (RTK), using various well-known procedures such as dead reckoning and triangulation. These methods are relatively costly and have not yet been proven as immediately applicable solutions in production vehicles. Often, digital maps need to be created to locate coordinates, so that there is an accurate representation of the road network with all the detailed information and features that would not be easy to locate using previous methods. Thus, localization and mapping are often performed simultaneously. One such approach is simultaneous localization and mapping (SLAM) \cite{b66,b65,b67}.
    \item SLAM (simultaneous localization and mapping): This process aligns the vehicle's positions with the digital map data, enhancing efforts to achieve accurate localization without using GNSS data. The aim of this approach is to use detection algorithms (EKF, LSD, FastSLAM 2.0) to create probabilistic models that evolve dynamically with the vehicle's movement in real time, providing a more realistic representation of the environment and the vehicle's position within it. These methods collectively address the question “Where am I?”, a fundamental aspect of safe and effective autonomous vehicle navigation \cite{b67,b65}.
    \item Mapping frameworks: Once the AV has found its position and destination, a mapping framework is created, which involves creating a detailed representation of the environment. Mapping systems rely on GNSS sensors with high-resolution maps based on geographic information systems (GIS), range and vision-based sensors, and cooperative approaches where local maps from individual vehicles are combined via vehicle-to-vehicle (V2V) communication. Maps can be categorized into offline and online maps. Offline maps, which are produced in advance, include digital and high-definition (HD) maps, which provide a detailed three-dimensional representation of the environment. In particular, HD maps are based on multiple layers (base, geometric, semantic, pre-knowledge, and real-time knowledge layers) providing specific information about the road environment. Online maps, developed through SLAM, address the challenge of navigating in unknown environments while simultaneously constructing the map and locating the vehicle \cite{b65}.
    \item Motion Planning: Route planning is the main function for determining and executing efficient vehicle trajectories. Route planning is divided into two main categories: global and local. Global route planning involves creating a macroscopic route from the vehicle's current position to its destination, using high-precision maps and taking into account known information about the road, such as route length and traffic volume. This type of planning, also known as static planning, does not take into account real-time environmental changes. In contrast, local route planning is dynamic, constantly adjusting the vehicle's trajectory based on real-time data from the vehicle's sensors. This allows for immediate response to obstacles and changing traffic conditions, ensuring collision-free navigation \cite{b67}.
    \item Multi-vehicle route planning: In addition to these individual vehicle processes, there is also simultaneous multi-vehicle route planning used in scenarios involving a fleet of autonomous vehicles. This method ensures coordinated movements and stability of the entire group, using advanced strategies such as aggregation, separation, and local behavior adjustments, with one vehicle acting as a leader to determine the path that the others will follow.
    \item Vehicle control: Vehicle control involves the precise management of the vehicle's movements based on the planned route. The so-called Drive-by-wire approach disconnects the mechanical parts from the driver's commands, replacing them with electronic controllers. This includes both longitudinal control, which manages the vehicle's speed through throttle and brake settings, and lateral control, which guides the vehicle's direction using the steering system. Effective motion control requires the system to perceive and anticipate potential hazards as far ahead as possible. Current technologies such as Adaptive Cruise Control (ACC) and Electronic Stability Programs (ESP) are an integral part of this process, maintaining safe distances between vehicles and ensuring stability during sudden accelerations \cite{Bordoloi2023,b71,b70}.
\end{enumerate}

These stages form a tightly coupled pipeline: detection feeds positioning and mapping; motion planning depends on accurate localization and updated environmental models; vehicle control translates planning decisions into physical motion through coordinated actuation.

Despite the progress in automation technologies, several challenges remain. These include handling sensor uncertainties in complex environments, maintaining accurate localization in urban areas with limited GNSS availability, ensuring real-time processing of large data volumes, and coordinating multiple vehicles in shared traffic environments. Addressing these challenges requires continued development of robust algorithms, high-performance computing units, and reliable communication infrastructures.

Examples of autonomous systems include:
\begin{itemize}
    \item \textbf{Autonomous parking:} Autonomous parking systems execute three stages---space search, autonomous parking maneuver, and vehicle return. GPS, IMU, LiDAR, radar, and cameras provide localization and obstacle detection; route planning accounts for turning radius, speed constraints, and space layout; and control algorithms ensure precise steering and acceleration throughout. Benefits include improved space utilization and reduction of fuel waste from manual searching \cite{b65,b66}.
    \item \textbf{Adaptive Cruise Control (ACC):} ACC maintains a safe following distance and desired speed without continuous driver input. It combines sensor-based detection, a central controller that decides throttle and braking commands, drive-by-wire actuators, and an HMI providing real-time traffic alerts. By compensating for delayed reactions or driver inattention, ACC reduces collision risk and supports the transition toward higher automation levels \cite{b71,b70,b72}.
\end{itemize}

Vehicle automation is formally classified according to SAE International Standard J3016, which defines six levels of driving automation (Levels 0--5), ranging from no automation to full automation. These levels provide a standard taxonomy widely adopted in the automotive industry and academic literature \cite{bSAE}. Integrating these advanced capabilities into vehicles requires a reliable and flexible architecture. The Automotive Open System Architecture (AUTOSAR) is the basic framework used by manufacturers to standardize and efficiently develop these systems.

Adaptive AUTOSAR extends the Classic AUTOSAR standard to support dynamic, software-updatable functions on high-performance computing platforms (perception, ADAS, OTA management), using POSIX-compliant OS interfaces and a runtime environment (RTE) that facilitates sensor processing, decision-making, and V2X communication \cite{Jiang2024,b50,b52}.

\subsubsection{AI technologies in the SDV context}

AI underpins every stage of the SDV automation pipeline, with different model families typically suited to different stages and operating constraints. The discussion below is intentionally kept at a general level, since precise benchmark figures and architecture choices vary across implementations and are evolving rapidly.

\textbf{CNNs for SDV perception:}
Convolutional architectures are widely used for camera-based object detection and for multi-modal fusion of camera and LiDAR data, and represent a substantial part of the perception literature for automated driving \cite{Liu2020}. In deployment, perception latency needs to remain compatible with the vehicle's planning cycle so that detections do not become stale by the time they reach the planning stack.

\textbf{Recurrent models for sequential data:}
LSTM/GRU-style recurrent encoders are commonly used for short-horizon trajectory prediction and for analysing time-series telemetry, for example toward early detection of developing faults from sensor trends \cite{b66,b74}. As with most learned components, achievable accuracy and prediction horizon depend heavily on the quality and representativeness of the available training data.

\textbf{Reinforcement learning for control and decision-making:}
Reinforcement-learning methods have been explored for tasks such as lane-change decisions and adaptive-cruise-control policy learning \cite{b66}. A central open issue is ensuring that learned policies respect hard safety requirements---such as collision avoidance---at every decision step; this is an active area of safe- and constrained-RL research rather than a solved problem.

\textbf{Lightweight and emerging approaches:}
Spiking neural networks (SNNs) paired with event-based (neuromorphic) vision sensors, and classical tree-based ensemble methods applied to on-board diagnostic data, are also being explored for SDVs because of their potential for low-power, low-latency inference on constrained ECUs. These approaches are comparatively less mature than CNN- and RNN-based methods and are not yet widely deployed in production vehicles \cite{b74,b75}.

\textbf{Hardware and deployment constraints:}
Running AI inference for perception, localization, and planning imposes a non-trivial computational and power load on SDV platforms \cite{Liu2020}. Automotive compute platforms therefore span a wide range, from higher-performance systems-on-chip with dedicated accelerators to low-power microcontrollers for resource-constrained ECUs, and techniques such as quantization, pruning, and knowledge distillation are commonly used to help fit models within these budgets.

No single AI technique satisfies every SDV requirement simultaneously: higher-capacity models tend to demand more compute and are harder to certify under safety standards such as ISO~26262, while lighter-weight methods are easier to deploy on constrained hardware but offer less representational capacity. SDV AI pipelines therefore typically combine several techniques across the automation stages described above, with the choice of technique and hardware tailored to the latency and power budget of each stage.

\section{Use Cases and Application Domains of SDVs}
This section presents key SDV use cases, including safety-critical systems, autonomous driving, connectivity, infotainment, fleet management, and Mobility as a Service.

\subsection{Safety-Critical Applications}
Safety-critical applications represent one of the most important domains of SDVs, as they directly affect road safety and passenger protection. These applications rely on real-time sensing, rapid data processing, and reliable communication to prevent accidents and mitigate hazardous situations \cite{Kukkala2018,Jiang2024}.

Typical safety-critical applications include collision avoidance systems, emergency braking, lane departure warning, blind spot detection, and driver monitoring systems \cite{Kukkala2018}. SDVs enhance these applications by enabling more centralized computing architectures and flexible software platforms that can improve perception, decision-making, and response performance \cite{Jiang2024}.

Software-defined architectures also allow safety algorithms to be updated remotely through OTA mechanisms, enabling vehicles to benefit from improved software functions without requiring hardware replacement \cite{b55}. In addition, integration with external communication systems such as vehicle-to-vehicle and vehicle-to-infrastructure communication enables vehicles to exchange hazard information with nearby vehicles and infrastructure, thereby improving situational awareness and cooperative road safety \cite{Chtourou2024}.

\subsection{Autonomous and Assisted Driving Applications}
Autonomous and assisted driving applications rely on sensor integration, AI algorithms, and high-performance computing platforms to support automated vehicle operation \cite{Liu2020,Jiang2024}.

Advanced Driver Assistance Systems (ADAS), such as adaptive cruise control, lane keeping assistance, automated parking, and traffic jam assist, represent early stages of vehicle autonomy. In fully autonomous systems, SDVs continuously process data from cameras, radar, and LiDAR sensors to perceive their surroundings and make driving decisions in real time \cite{Liu2020}.

Software-defined platforms enable the continuous improvement of autonomous driving capabilities through remote software updates and modular deployment of new algorithms. This flexibility allows manufacturers to refine perception models, enhance driving strategies, and introduce new features throughout the vehicle lifecycle \cite{Jiang2024}.

\subsection{Connectivity and Cooperative Driving Applications}
Connectivity and cooperative driving represent key application domains of SDVs, enabling vehicles to interact with each other and with surrounding infrastructure through Vehicle-to-Everything (V2X) communication technologies \cite{Abdelkader2021,Malik2021}.

Through Vehicle-to-Vehicle (V2V) and Vehicle-to-Infrastructure (V2I) communication, SDVs can exchange real-time information such as vehicle speed, position, and trajectory. This enables cooperative awareness, traffic coordination, and hazard notification, significantly improving safety and traffic efficiency \cite{Abdelkader2021,Malik2021}.

Cooperative driving applications include platooning, intersection management, emergency vehicle prioritization, and collaborative route optimization. These applications require low-latency communication networks, such as 5G and beyond, to ensure reliable data exchange \cite{Malik2021,Hakak2022}.

Software-Defined Networking (SDN) approaches further enhance cooperative driving by enabling centralized traffic management and dynamic network configuration, allowing transportation systems to adapt to changing traffic conditions.

\subsection{Infotainment Systems}
The term ``infotainment'' combines ``information'' and ``entertainment,'' referring to the integration of information and entertainment functions into a single system. In SDVs, infotainment systems have evolved from simple music players into complex platforms combining navigation, communication, and entertainment services \cite{b29,b47}. They act as central hubs connecting the vehicle to the outside world via mobile and Wi-Fi networks, enabling two-way communication with smartphones and integration with ADAS and safety systems \cite{b47,b29}. Machine learning allows these systems to adapt to each driver's preferences, providing a personalized experience \cite{b47}. Core IVI functions include:
\begin{itemize}
    \item \textbf{Navigation}: Often combined with real-time traffic data.
    \item \textbf{Entertainment}: Audio and video playback, internet streaming, and games.
    \item \textbf{Communication}: Hands-free calling, voice recognition, and messaging.
    \item \textbf{Connectivity}: Wi-Fi, Bluetooth, and smartphone integration (Apple CarPlay, Android Auto).
\end{itemize}
These functions are expected to work seamlessly without interfering with the primary purpose of the vehicle, which is safe transportation. This requires careful consideration of user interface design and system architecture to avoid distracting the driver while providing rich functionality to passengers. \cite{b29}

IVI systems face challenges distinct from consumer electronics: longer product lifecycles (5--10$\times$ consumer devices) with implications for update management and hardware robustness; mandatory safety isolation to prevent third-party apps from interfering with critical vehicle functions; and platform fragmentation across manufacturers despite standardization efforts (GENIVI Alliance) \cite{b29,b47}. The dominant platforms are:
\begin{itemize}
    \item \textbf{AAOS}: A dedicated in-car OS running directly on the vehicle hardware (unlike Android Auto, which mirrors a phone). Supports both Google and third-party apps via a developer SDK, with safety isolation preventing interference with critical vehicle functions.
    \item \textbf{QNX}: BlackBerry's Unix-like RTOS known for reliability and security. Its microkernel design runs processes in isolation, ensuring stability under software faults, and extends to ADAS and autonomous driving functions.
    \item \textbf{AGL}: Linux Foundation open-source platform for infotainment, instrument cluster, and telematics, supporting navigation, V2X connectivity, and extensive manufacturer customization.
\end{itemize}

As vehicles become more autonomous, infotainment will evolve into a mobile hub for teleconferencing, VR, and commerce, enabled by 5G streaming, V2X integration, and cloud offloading for NLP and voice recognition \cite{b47}.

\subsection{Software-Driven Fleet Management}
SDV solutions transform fleet vehicles into mobile data platforms, enabling real-time GPS tracking, preventive maintenance alerts, and OTA updates \cite{b74,b75}. Knowing the exact location and status of each vehicle at all times can dramatically improve efficiency, particularly in supply chain management where time-sensitive deliveries are critical \cite{b74,b75}.

Modern real-time monitoring relies on advanced GPS systems and sensors that record location, engine status, fuel consumption, and other parameters at millisecond intervals. Data is transmitted via 5G to cloud analytics platforms where AI/ML algorithms identify patterns, predict maintenance needs, and optimize routes \cite{b74,b75}. Processed insights are presented via dashboards enabling rerouting decisions and predictive maintenance scheduling.

In practice, commercial fleet-telematics platforms combine real-time GPS tracking, driver-behavior analysis, preventive-maintenance alerts, and battery monitoring for electric fleets, drawing on cloud-based analytics and OTA updates to reduce operational costs and support safety compliance \cite{b74,b75}.

\subsection{Mobility as a Service (MaaS)}
Mobility as a Service (MaaS) is a transformative concept in transportation that integrates multiple modes of transportation into a seamless, integrated service. It provides a user-centric solution to transportation needs, optimizing time, cost, and sustainability through a flexible digital platform. On the other hand, SDVs offer dynamic capabilities that are perfectly aligned with the objectives of MaaS.

At the core of SDVs is their ability to be continuously upgraded, optimized, and controlled through software updates, making them highly adaptable to changing mobility demands. When combined with MaaS, SDVs offer increased flexibility and operational efficiency for transportation networks. Here is how SDVs can be integrated into MaaS platforms:
\begin{itemize}
    \item Real-time adaptability: SDVs can be reconfigured in real time to meet the diverse needs of users. For example, SDVs can switch between ride-sharing services, public transport integration, or even real-time on-demand goods delivery services. MaaS platforms can leverage this capability for dynamic allocation of SDVs based on factors such as time of day, traffic, or user preferences. \cite{b74,b75}
    \item Personalized mobility: MaaS platforms aim to provide personalized travel experiences, and SDVs reinforce this goal through software-based, customized vehicle settings. From in-car entertainment to the vehicle's interior environment, SDVs can customize features such as seat preferences, climate control, and route suggestions, improving the user experience based on data from the MaaS platform.
    \item Vehicle as a service (VaaS): MaaS platforms could enable SDVs to be reconfigured for different use cases, such as taxis, autonomous buses, or personal routes, depending on user demand. SDVs could be autonomously adapted to serve different customer segments, ensuring that transportation is provided efficiently based on specific needs, from comfort to cost-effective shared rides. \cite{b74,b75}
\end{itemize}

SDVs serve as first- and last-mile solutions in congested urban areas, complement public transport during peak hours, and adapt to shared or individual service modes during off-peak periods \cite{b74,b75}. Electric SDVs deployed in shared mobility models further reduce emissions at scale, while route optimization algorithms minimize idle time and fuel consumption \cite{b74,b75}.

\subsection{Applications of AI in the context of SDVs}

AI provides autonomous vehicles with capabilities that enable them to operate more safely and efficiently. Two representative application areas illustrate this role:

\textbf{\textit{Autonomous driving}}: Learning-based methods are applied across the perception, localization, mapping, and motion-control stages of the automation pipeline, for instance to support object detection or trajectory and route planning \cite{b66,b67}. The specific techniques used, and their relative maturity, vary considerably across these stages.

\textbf{\textit{Predictive Maintenance}}: AI-assisted analysis of sensor data such as engine temperature, vibration, and tire pressure can help flag early signs of component degradation, complementing conventional rule-based diagnostic methods \cite{b74,b75}. Across both areas, AI functions as a supporting tool for perception, decision-making, and diagnostics rather than a fully autonomous replacement for established engineering and validation processes.

\subsection{Summary and Key Observations}
The use cases span safety-critical systems, autonomous driving, cooperative connectivity, infotainment, fleet management, and MaaS, collectively demonstrating the need for scalable, secure, and interoperable SDV architectures. Table~\ref{tab:sdv_use_cases} summarizes these domains with their enabling technologies, communication needs, and latency requirements.

\begin{table*}[!t]
\centering
\caption{Classification of SDV Use Cases}
\label{tab:sdv_use_cases}
\renewcommand{\arraystretch}{1.1}
\setlength{\tabcolsep}{4pt}
\footnotesize
\begin{tabular*}{\textwidth}{@{\extracolsep{\fill}}|p{3.0cm}|p{2.5cm}|p{4.5cm}|p{3.0cm}|p{3.0cm}|}
\hline
\textbf{Use Case} &
\textbf{Domain} &
\textbf{Key Technologies} &
\textbf{Communication Requirements} &
\textbf{Latency Requirements} \\
\hline

Collision Avoidance &
Safety-Critical &
Sensors, ADAS, AI Processing &
V2V, In-Vehicle Networks &
Ultra-Low \\
\hline

Autonomous Driving &
Automation &
AI, LiDAR, Cameras, Edge Computing &
High-Bandwidth V2X &
Ultra-Low \\
\hline

Cooperative Driving &
Connectivity &
V2X, SDN, Cloud Services &
V2V, V2I &
Very Low \\
\hline

Infotainment Systems &
User Services &
Cloud Services, OS Platforms &
Wi-Fi, Cellular &
Medium \\
\hline

Fleet Management &
Mobility Services &
IoT, GPS, Cloud Analytics &
Cellular, Cloud &
Medium \\
\hline

MaaS &
Smart Mobility &
Cloud Platforms, AI Optimization &
V2X, Cloud Integration &
Medium \\
\hline

\end{tabular*}
\end{table*}

\section{SDVs BEYOND THE VEHICLE: THE SOFTWARE-DEFINED INTERNET OF VEHICLES}
Although SDVs primarily focus on software-centric vehicle architectures, centralized computing platforms, and continuous software lifecycle management, many of their advanced functionalities depend on communication and computing resources beyond the vehicle itself. Functions such as cooperative driving, V2X communication, cloud-assisted perception, fleet management, and over-the-air software updates require seamless interaction with roadside infrastructure, edge nodes, and cloud platforms. Consequently, the software-defined paradigm is increasingly extending from the vehicle domain to the broader vehicular ecosystem.

This evolution has given rise to the Software-Defined Internet of Vehicles (SDIoV), which combines software-defined networking principles with vehicular communication, edge computing, and cloud infrastructures to enable flexible, programmable, and scalable transportation systems. Rather than representing a separate research direction, SDIoV can be viewed as a natural extension of the SDV paradigm, expanding software-defined control and intelligence from individual vehicles to end-to-end vehicular environments. The following section reviews the key architectural principles, enabling technologies, and operational benefits of SDIoV systems.

\subsection{Overview of the Internet of Vehicles (IoV)}
The Internet of Vehicles (IoV) extends IoT to transportation, connecting vehicles, infrastructure, and pedestrians via V2V, V2I, and V2P communication to improve road safety, traffic management, and services \cite{b104}. Building on VANETs, IoV integrates 4G/5G, Wi-Fi, DSRC, and DSA for multi-layer data exchange and real-time services. Key challenges include: scalability under high vehicle mobility and concurrent communications; heterogeneous QoS requirements (ultra-low latency for safety vs. higher tolerance for infotainment); seamless handoff across heterogeneous radio technologies; cybersecurity and privacy of connected vehicles \cite{b114}; and dynamic bandwidth and compute resource management \cite{b104}.

\subsection{Software-Defined Networking (SDN)}
SDN decouples control and data planes, centralizes network management, and enables programmable, software-driven network behavior for dynamic adaptation and efficiency improvements.

The SDN architecture is structured in three layers, each with a distinct role \cite{b104,Cardona2020}:

\begin{enumerate}
    \item \textbf{Data Plane}: The data plane, also known as the infrastructure layer, is responsible for forwarding data packets based on rules provided by the control plane. The networking devices at this level follow the specifications set by the SDN controller without making their own decisions about routing or traffic management. This simplifies device operation, reducing complexity and improving packet forwarding efficiency.
    \item \textbf{Control Plane}: The control plane in SDN is centralized in the SDN controller. This layer makes routing decisions, creates policies, and dictates how the data plane should handle traffic. The SDN controller has a global view of the entire network, which allows it to optimize routing paths, improve traffic management, and ensure that network resources are used efficiently. By decoupling this functionality from individual network devices, SDN creates a more flexible network capable of real-time adjustments.
    \item \textbf{Application Layer}: The application layer interfaces with the control layer, allowing administrators to create high-level network policies and services. Applications such as firewalls, load balancers, and intrusion detection systems reside at this layer, using central control to enforce policies across the network. By abstracting away the network hardware, the application layer allows customized applications to control network behavior based on specific organizational needs, making SDN highly adaptable.
\end{enumerate}

SDN controllers are the central elements managing the network. Representative open-source examples include NOX/POX (Python-based, OpenFlow-compliant), FloodLight (Java-based, commercial-grade ease of use), RYU (NTT, flexible multi-protocol API), OpenDaylight (supports both OpenFlow and legacy devices), and ONOS (distributed, suitable for large-scale deployments) \cite{Cardona2020}. Controller selection depends on scale, performance requirements, and customization needs.

\subsection{SDIoV Solution Framework}
SDIoV applies SDN principles to vehicular networks, addressing the scalability and dynamic management limitations of traditional VANETs. By separating network control from physical hardware, SDIoV enables centralized, programmable communication management and faster coordination among vehicles, infrastructure, and pedestrians \cite{Cardona2020}. Recent approaches further extend SDIoV with intelligent orchestration of controller placement and task offloading to improve resilience and load balancing under dynamic vehicular conditions \cite{MoToBFT,Haq25}. This section examines three foundational architectures: the basic SDIoV framework, and its extensions incorporating edge and fog computing \cite{Cardona2020,Haq25,Nke23}.

\subsubsection{Basic SDIoV Architecture}
In the basic SD-IoV (Software-Defined Internet of Vehicles) architecture, the main elements include: logical controllers (SDN controllers), the network of SDN switches, wireless access infrastructures (BSs, RSUs, etc.), vehicles (with SDN capability), and data and control paths. Each of these components plays an important role in supporting the network connectivity and management of the IoV.

\begin{itemize}
    \item SDN logic controllers: SDN logic controllers are central software programs that run on servers, either in the cloud or locally, depending on how quickly they need to respond and how easily they need to adapt. These controllers have a comprehensive view of the network and are responsible for various tasks, such as network management, rule creation, and resource allocation. In addition, they perform more complex functions, such as data preprocessing and network analysis. The term ``logical'' means that controllers can be distributed and work together in a hierarchical manner \cite{Cardona2020}.
    \item SDN switch network: This network consists of special switches that support SDN and manage the flow of data between the Internet and vehicles. When data arrives at the network, the switches forward it using predefined rules. If there are no suitable rules, the data is sent to the SDN controller, which decides the correct route and sends the instructions back to the switches. This network can be scaled up to a large extent, as Google does with its SD-WAN \cite{Cardona2020}.
    \item SDN wireless access infrastructure: These infrastructures include various wireless access devices, such as Wi-Fi access points (APs), mobile base stations (BSs), roadside units (RSUs), and coordinators for Dynamic Spectrum Access (DSA). These infrastructures provide the wireless connectivity required for vehicles and are also controlled by the SDN for managing tasks such as power control, channel assignment, and resource allocation \cite{Cardona2020}.
    \item SDN-enabled vehicles (OBUs): Vehicles equipped with OBUs are the end users of the network and, in some cases, also act as relays for other vehicle-to-vehicle (V2V) communications. OBUs can support various SDN functions, including packet forwarding and interface selection, depending on their hardware capabilities \cite{Cardona2020}.
    \item Data Path: Data paths in SDIoV include both wired and wireless connections. Wired connections connect the Internet, SDN switch networks, and wireless access infrastructures, while wireless connections facilitate communication between SDN-enabled vehicles. These paths use different technologies without changing the lower layers (PHY and MAC), focusing on logical wireless control paths for connecting vehicles and controllers \cite{Cardona2020}.
    \item Control Path: The control path is necessary for SDN controllers to manage the system in real time and receive feedback from other elements. It ensures that controllers can send instructions to SDN switches, wireless infrastructures, and vehicles. Wired control paths connect controllers and switches, while wireless control paths are critical for managing communication between vehicles and controllers. This often requires special channels and protocols for sending commands \cite{Cardona2020}.
\end{itemize}
The individual layers of this SDIoV architecture (based on SDN logic) are the following:

\begin{enumerate}
    \item Application and Control Layers: These layers are located in data centers, either in the cloud or on local servers. The application layer hosts network services such as traffic monitoring and data analysis. It works with the control layer to enforce rules related to data flow management and network resource optimization. This collaboration is similar to the operation of software and operating systems on computers \cite{Cardona2020}.
    \item Upper Data Layer: This layer includes SDN switches and wireless access infrastructure, such as Wi-Fi access points, mobile base stations, and roadside units. This infrastructure is responsible for the physical transmission of data over wired networks. They use the OpenFlow protocol, which enables efficient packet forwarding management, power control, and other advanced features such as traffic activation and monitoring \cite{Cardona2020}.
    \item Lower Data Layer: Consists of SDN-enabled vehicles, also known as OBUs (On-Board Units). This layer is responsible for wireless data exchange between vehicles. OBUs communicate with SDN switches and wireless access infrastructure and require customized versions of the OpenFlow protocol to handle specific vehicle requirements such as interface management and traffic management \cite{Cardona2020}.
    \item Awareness Layer: This layer collects information about the state of the network. It provides critical information to other layers and performs enhancements such as data reduction and privacy protection to improve network efficiency \cite{Cardona2020}.
\end{enumerate}

\subsubsection{Edge Computing-Enabled SDIoV (EC-SDIoV)}
Intelligent edge computing plays an important role in modern networks, bringing data processing, storage, and services closer to the devices that generate the data, such as vehicles in the Internet of Vehicles (IoV). This helps reduce the amount of data sent to central servers, thereby reducing network traffic and improving speed and performance \cite{Haq25,b94}. In the IoV, intelligent edge computing offers many other benefits:
\begin{enumerate}
    \item Real-time decision making: Vehicles and nearby units (RSUs) can make immediate decisions about important functions such as routing and location tracking \cite{Haq25,b94}.
    \item Data sharing: By combining edge computing with SDN technologies, vehicles can share data quickly and efficiently, enhancing communication and collaboration \cite{Haq25}.
    \item Distributed learning: Vehicles and RSUs work together to learn and execute decisions, connecting to the cloud for additional support, thereby improving performance and adaptability \cite{Haq25,bN5}.
    \item Service offloading: Edge computing allows vehicles to ``offload'' some of their computing needs to edge units, facilitating functions such as autonomous navigation and accurate mapping of the environment \cite{Haq25,b94}.
\end{enumerate}
The Edge Computing-Enabled Software-Defined Internet of Vehicles (EC-SDIoV) architecture is designed to support the increased demands of modern vehicle networks by integrating edge computing and SDN. The architecture is based on communication between two basic levels: the data level and the control level \cite{Haq25,b94}.

More specifically, the data layer consists of fixed and mobile computing nodes. Fixed nodes, such as Road-Side Units (RSUs), are located at fixed points and are directly connected to the network control center. Mobile nodes are the vehicles themselves, which have powerful computing resources. These vehicles can process data locally and participate in the transmission of information while in motion. The key innovation offered by SDN logic here is that data is transferred and processed independently of control. This helps improve resource management and reduces communication delays. For example, vehicles with excess computing power can ``lease'' it to other users on the network, helping to manage resources in real time \cite{Haq25}.

At the same time, at the control level, there is the SDN controller, which is responsible for managing and supervising the entire network. This controller continuously monitors resource usage, the status of communication channels, service requirements, and network topology. Based on this data, it makes decisions about data routing and the allocation of computing resources. To manage vehicle mobility, the SDN controller works with RSUs, which act as intermediate nodes. This strategy allows the network to remain flexible and avoid delays that can be caused by changes in topology. Recent work has addressed the joint optimisation of controller placement and task offloading in vehicular edge SDN using gradient-driven orchestration with Byzantine Fault Tolerant (BFT) guarantees, demonstrating improved resilience and load balancing under dynamic vehicular conditions \cite{MoToBFT}.

EC-SDIoV's decentralized design improves reliability (local processing survives central server failures), reduces congestion (only necessary data is transmitted), enhances privacy, and enables collaborative resource sharing where high-compute vehicles assist those with limited resources. Overall, EC-SDIoV is among the most promising developments for connected and intelligent mobility. It is worth noting that the European Telecommunications Standards Institute (ETSI) has developed a standardized framework for Multi-Access Edge Computing (MEC), which provides the normative specifications for deploying computing capabilities at the network edge in vehicular environments. ETSI MEC defines APIs, service interfaces, and reference architectures that are widely adopted by network operators and automotive OEMs, making it a critical enabler for the EC-SDIoV architecture described in this section \cite{bETSI}.

\subsubsection{Fog Computing-Enabled SDIoV (FC-SDIoV)}
Fog Computing has emerged as a transformative paradigm in the field of the Internet of Vehicles (SDIoV), significantly improving the capabilities of vehicle networks. More specifically, in the context of SDIoV, it refers to a decentralized model where computing resources (processing, storage, and networking) are distributed closer to the data source, such as vehicles or roadside units (RSUs) \cite{Nke23,Haq25}.

Regarding the relationship between Fog Computing and Edge Computing in SDIoV, fog computing is often considered an extension of edge computing, but with a more distributed character. While edge computing focuses on devices such as the vehicles themselves or sensors, fog computing introduces an additional layer of processing between the edge and the cloud, enabling a hierarchical, multi-level approach to resource management in SDIoV \cite{Nke23}.

As discussed in the previous section, edge computing serves fast, local decisions and processes data on individual devices. In fog computing, the logic ensures that the system as a whole can handle more complex tasks, manage resources efficiently, and adapt to the challenges posed by mobility and large-scale data production in the IoV \cite{Nke23,Haq25}.

Fog computing enhances IoV efficiency through several mechanisms: flexible resource allocation across multiple edge devices; reduced latency by processing data close to its source; optimized bandwidth use by retaining most data at fog nodes and transmitting only what is necessary to the cloud; continuous mobility support through handoff between RSUs and local edge servers; and improved network scalability via distributed load management that avoids bottlenecks at the central infrastructure.

The fog computing architecture in SDIoV is structured as a multi-level system that optimizes resource management and service provision. It consists of: vehicles, fog servers, the cloud, and the communication layer that connects all these elements together.

\begin{itemize}
    \item Edge vehicles: This layer includes individual vehicles equipped with various sensors and communication devices. These vehicles serve as both data sources and processing units, collecting information about their environment through GPS, lidar, and radar technologies. They communicate with each other (V2V) and with RSUs (Vehicle-to-Roadside, V2R) to share critical information. This interaction not only enhances safety but also promotes a collaborative environment where vehicles can respond effectively to real-time conditions \cite{b104}.
    \item Fog Nodes: These are computing nodes located close to data sources, such as sensors, IoT devices, or vehicles, and act as an intermediate link. They typically distribute power and storage resources locally, enabling immediate data processing, reducing latency and load on the central cloud. These nodes can be located in infrastructure such as servers placed at strategic points in the network and are used to enable fast, real-time responses in applications such as traffic management. An important piece of information worth mentioning is that sometimes RSUs can function as fog nodes in such a network. Located near roads and areas with heavy traffic, RSUs collect and analyze data from nearby infrastructure. This process enhances the effectiveness of applications such as hazard warning systems. This local processing makes them ideal fog nodes, as they can provide services faster and more efficiently than a central cloud server \cite{Nke23,Haq25}.
    \item Cloud infrastructure: At the top tier, cloud servers offer extensive computing and storage capabilities. Although they generally involve higher latency, they are essential for complex data analysis and long-term storage. This layer can effectively manage large-scale applications such as traffic management systems and data analysis for smart city planning \cite{Nke23}.
    \item Communication layer: This is necessary to establish connections between the cloud, fog nodes, and end devices, enabling the smooth transfer of data, control signals, and task offloading to different layers. It uses a range of protocols and technologies, including high-speed wireless communication standards such as 5G, 4G, and LTE, which are vital for mobile devices and vehicle networks. It also uses Wi-Fi and Bluetooth for local communication between edge devices and nearby fog nodes, and Ethernet to facilitate wired communication between fog nodes and cloud data centers \cite{b104}.
\end{itemize}

\section{SDV Challenges}
The automotive industry is undergoing a period of rapid transition, with technological innovations reshaping the landscape. At the same time, critical challenges are emerging across five key areas: big data management, cybersecurity and data privacy, energy efficiency, interoperability and standardisation, and functional safety. This section analyses these challenges and concludes with emerging future research directions.

\subsection{Big Data Handling}
SDVs generate and manage vast amounts of data from sensors, cameras, and communication systems, with a modern autonomous vehicle capable of producing, depending on its sensor configuration, on the order of several terabytes per driving hour \cite{Liu2020}. Three distinct challenges arise from this volume and velocity. First, \textbf{real-time processing}: unlike generic IoT pipelines where a missed deadline degrades service quality, SDV perception, localization, and planning operate under \emph{safety-critical} deadlines---a processing overrun can trigger fail-safe interventions or, worse, allow unsafe decisions to propagate. Multi-modal sensor fusion across cameras, LiDAR, and radar---each running at different frame rates---must be temporally synchronized and processed on a thermally and power-constrained mobile platform that cannot simply defer computation to the cloud when network connectivity is absent \cite{Liu2020,Haq25}. Second, \textbf{storage and bandwidth}: raw sensor logs cannot be discarded after each trip; regulatory liability frameworks, safety investigations, and model retraining pipelines all require event data retention, yet wireless bandwidth during driving is insufficient to upload continuous high-rate streams. OTA update delivery to large fleets introduces an additional battery and cellular bandwidth burden that must be managed without competing with safety-critical V2X communication channels \cite{Haq25,b55}. Third, \textbf{data quality}: the long-tail problem is uniquely severe in SDV deployments---rare and safety-critical scenarios such as sensor degradation, adversarial road markings, or unusual obstacle geometries are precisely the cases that matter most for safety yet are most underrepresented in training data. Multi-modal annotation across synchronized LiDAR, camera, and radar streams is substantially more complex and costly than single-sensor image labeling, making it difficult to maintain annotation quality at the scale and geographic diversity that production SDV models require \cite{Liu2020}.

\subsection{Cybersecurity and Data Privacy}
SDVs expose multiple attack surfaces that are distinct from those of conventional vehicles: V2V/V2I communication channels, Wi-Fi/Bluetooth/5G interfaces, infotainment systems, telematics units, and OTA update pipelines. Demonstrations of remote vehicle exploitation, in which researchers have reached safety-critical functions through connectivity and infotainment interfaces, illustrate the severity of these vulnerabilities \cite{b113}. Compromised OTA channels could inject malware into safety-critical systems, while adversarial attacks on AI perception (e.g., manipulated traffic signs or sensor readings) can induce unsafe decisions \cite{bN3}. V2X integrity requires strong authentication and real-time anomaly detection to prevent communication-level interference \cite{bN3}.

Beyond active attack surfaces, \textbf{data privacy and ownership} present a distinct regulatory challenge. Connected vehicles continuously generate personal and geolocation data; under GDPR, this data requires explicit consent and purpose limitation, yet ownership is contested among the driver, the manufacturer, and third-party service providers \cite{bN3}. A comprehensive analysis of SDV cybersecurity and privacy challenges identifies that software-defined architectures introduce new attack classes including API vulnerabilities, third-party software risks, and supply-chain threats \cite{bN3}. Multi-layered defence strategies integrating both in-vehicle and cloud-based security mechanisms are required to address this expanded attack surface. A security-by-design framework for OTA updates addresses key gaps in automotive cybersecurity practice \cite{bN2}, and compliance with UNECE Regulation No.~155 (Cybersecurity Management System) and No.~156 (Software Update Management System) is now mandatory for vehicle type approval in participating markets \cite{bR155,bN4}.

\subsection{Energy Efficiency}
The AI inference stack, V2X communication, and always-on sensor processing collectively impose energy demands on SDV platforms that have no equivalent in conventional vehicles. Deep learning models for perception and sensor fusion---consuming, illustratively, on the order of tens to a few hundred watts depending on the specific system-on-chip and workload, well below the multi-kilowatt power draw of the vehicle's full sensing, computing, and communication stack---compete for power with continuous 5G/V2X radio operation, active cooling of high-performance SoCs, and electromechanical actuators \cite{b107}. For battery electric SDVs, this is especially critical: every watt consumed by computing and communication directly reduces driving range.

Three specific challenges arise. First, \textbf{inference efficiency}: large perception models cannot simply be ported from data-centre training to in-vehicle deployment; quantization (INT8/INT4), pruning, and knowledge distillation are required to bring latency and power within automotive budgets, but these techniques can degrade accuracy in edge cases such as adverse weather or occlusion \cite{Liu2020}. Second, \textbf{communication energy}: duty-cycling V2X radios and applying intelligent offloading policies---transmitting only safety-critical data continuously, buffering non-critical telemetry---can reduce communication energy without violating latency requirements for cooperative driving \cite{b107,Haq25}. Third, \textbf{OTA update energy}: large firmware updates transmitted to fleets during peak hours create significant load on both vehicle batteries and cellular infrastructure; scheduling and differential update mechanisms that minimise transmission volume are an active area of standardisation \cite{bN4}.

\subsection{Interoperability and Standardisation}

The proliferation of heterogeneous hardware platforms, operating systems, middleware solutions, and communication protocols across vehicle manufacturers and suppliers creates significant interoperability challenges for SDV ecosystems. The absence of universally accepted standards for software interfaces, update mechanisms, and service-oriented communication increases integration complexity and limits the portability of software components across platforms.

Current standardisation activity spans four interconnected domains. At the \textbf{in-vehicle software layer}, AUTOSAR Adaptive provides a POSIX-compliant execution environment and service-oriented communication model that enables hardware--software decoupling and OTA-deployable functionality across heterogeneous ECU platforms \cite{b50,b52}. At the \textbf{network edge}, ETSI MEC defines normative APIs, service interfaces, and reference architectures for deploying compute capabilities at the vehicular network edge, and is widely adopted by network operators and automotive OEMs as the basis for edge-assisted SDV and SDIoV deployments \cite{bETSI}. At the \textbf{V2X communication layer}, a fragmented standardisation landscape persists between IEEE~802.11p/WAVE (DSRC-based) and 3GPP C-V2X technologies (including LTE-V2X and NR-V2X), with ongoing efforts toward coexistence and convergence among alternative vehicular communication frameworks \cite{b80211p,b3GPPV2X,Haq25}. At the \textbf{cybersecurity and update management layer}, UNECE Regulations No.~155 and No.~156 now mandate certified Cybersecurity Management Systems and Software Update Management Systems for vehicle type approval in participating markets, directly shaping how OTA security must be implemented across the SDV lifecycle \cite{bR155,bR156,b21434}.

Recent ecosystem initiatives further complement formal standardisation efforts by promoting interoperable software platforms and common development frameworks for SDVs. SOAFEE (Scalable Open Architecture for Embedded Edge) advocates cloud-native software development, containerized workloads, virtualization, and hardware-independent deployment across automotive platforms \cite{bSOAFEE}. In parallel, the Connected Vehicle Systems Alliance (COVESA) coordinates industry-wide efforts toward software interoperability and data portability across the automotive ecosystem \cite{bCOVESA}. Among its most prominent initiatives, the Vehicle Signal Specification (VSS) provides a common vehicle data model that facilitates interoperable access to vehicle signals across OEMs, suppliers, and cloud services \cite{bVSS}. Furthermore, the Eclipse Software Defined Vehicle (SDV) Working Group promotes open-source technologies and collaborative development frameworks for software-defined mobility \cite{bEclipseSDV}, while the S-CORE (Safe Open Vehicle Core) project focuses on the development of a safety-oriented open-source software stack suitable for future SDV platforms \cite{bSCORE}. Collectively, these initiatives demonstrate an industry-wide effort toward improving interoperability, portability, software reuse, and ecosystem-wide collaboration across heterogeneous SDV deployments.

Despite this activity, full harmonisation across the automotive supply chain remains an open challenge. As surveyed in \cite{Haq25}, the coexistence of multiple SDN controller frameworks, edge computing APIs, middleware implementations, and V2X radio access technologies introduces integration overhead that existing standards do not yet fully resolve. Current standardisation efforts largely address individual layers of the SDV stack; however, the integration of vehicle software platforms, communication infrastructures, edge-cloud resources, and regulatory frameworks into a unified software-defined mobility architecture remains incomplete. Future research should therefore focus on cross-layer orchestration mechanisms, interoperable service interfaces, and common data models capable of supporting seamless operation across vehicle, edge, cloud, and V2X environments.

\subsection{Functional Safety and System Complexity}
ISO~26262 was designed for vehicles with static, hardware-bound functions certified at production time. SDVs violate this assumption at two levels. First, \textbf{OTA-induced lifecycle change}: a software update pushed months after type approval can alter safety-relevant behaviour in braking, steering, or ADAS functions; the existing standard does not provide a normative process for re-certifying safety claims after field deployment, leaving a compliance gap that UNECE Regulation No.~156 partially addresses through Software Update Management Systems but does not fully close \cite{bN2,bN4,b53}. Second, \textbf{shared-hardware interference}: centralised HPCU architectures run safety-critical functions (e.g., AEB, lane keeping) and non-critical services (e.g., infotainment, OTA management) on the same compute platform. Achieving ASIL-B or ASIL-D compliance in this context requires spatial and temporal partitioning---hypervisor-enforced memory isolation, real-time scheduling guarantees, and freedom-from-interference proofs---that add significant engineering complexity and limit the flexibility that software-defined design is meant to provide \cite{b53,Jiang2024,bN2}.

A further emerging challenge is \textbf{AI functional safety}: deep learning perception models used in ADAS and autonomous driving functions do not admit formal verification by conventional methods. ISO~21448 (Safety of the Intended Functionality, SOTIF) was developed specifically to address performance limitations and misuse in safety-related intended functionality, making it the normative framework most directly applicable to AI-based perception and decision systems in SDVs. Establishing safety cases for neural networks under distribution shift---when a model encounters road conditions absent from its training data---remains an open research problem with direct regulatory implications as ISO~21448 and successor standards begin to address AI-based driving functions \cite{Liu2020}.

\subsection{Future Research Directions}

\textbf{Digital twins for OTA validation and predictive maintenance.}
Digital twins---continuously synchronised virtual replicas of physical vehicles---offer a principled solution to the OTA certification gap identified above. A software update can be validated against the twin under simulated edge-case conditions before deployment to the physical fleet, dramatically reducing the safety-validation overhead of FOTA cycles. Extending digital twins to predictive maintenance enables failure prognosis from real-time sensor telemetry without requiring physical inspection, reducing fleet downtime and operational cost. Integrating digital twins into the SDV OTA lifecycle is an active standardisation and research direction.

\textbf{Federated learning for privacy-preserving AI improvement.}
Centralised retraining of SDV perception and predictive maintenance models requires uploading raw driving data to cloud infrastructure, raising privacy and bandwidth concerns at scale. Federated learning (FL) enables vehicles to train local model updates on-device and share only gradient information with a central aggregator, preserving data locality while enabling collaborative improvement \cite{bN5}. Key open challenges specific to the SDV context include handling highly non-IID data distributions across geographically diverse fleets, managing unreliable vehicle connectivity during aggregation rounds, and certifying that FL-updated models satisfy functional safety requirements before deployment.

\textbf{Proactive cybersecurity at the edge.}
As the SDV attack surface expands, reactive perimeter defences are insufficient. Proactive approaches deploy lightweight anomaly detection and intrusion detection systems (IDS) directly on vehicle ECUs and RSUs, using TinyML models compressed for microcontroller deployment. Reinforcement learning-based adaptive security policies can respond to novel attack patterns in real time without requiring cloud connectivity, which is essential for vehicles operating in areas with intermittent network coverage. Network failure detection and recovery---including controller failover in SDIoV architectures---is a related open problem: SDN controller failure in a vehicular network can degrade safety-critical V2X services, and resilient multi-controller placement strategies that jointly optimise placement and task offloading under Byzantine fault conditions represent an active research frontier \cite{b99,MoToBFT}.

\textbf{AI-defined vehicles and LLM-driven interfaces.}
Large language models (LLMs) are beginning to be integrated into the V2X communication layer as natural-language interfaces for in-vehicle assistants and cooperative driving coordination \cite{bN7}. Looking further ahead, AI-Defined Vehicles (AIDVs)---where continuously learning AI agents govern vehicle behaviour and adapt to driver preferences, traffic conditions, and regulatory environments without explicit reprogramming---represent the next evolution beyond SDVs \cite{bN8}. This paradigm introduces fundamental questions about auditability, liability, and the limits of human override that go beyond current SDV engineering frameworks and will require new regulatory and safety standards.

\section{Conclusions and Future Advancements}

This work provides a review of SDVs, focusing on the evolution of SDVs and their key technical challenges. The paper adopts the five-domains classification, namely, hardware and sensing, electrical/electronic architecture, software and service-oriented frameworks, automation, AI, and cloud/distributed infrastructure. As shown in the paper, this five-domain framework can provide a unified approach for capturing the complexity and interconnectivity of contemporary SDV systems.

From a technological perspective, SDVs represent the latest generation of vehicles that are shifting from hardware-oriented and ECU-based architectures toward highly integrated centralized platforms and software, enabling constant evolution of vehicles as cyber-physical systems. Specifically, as far as the hardware and architecture design is concerned, the transition toward more powerful central computers that perform various vehicle functionalities has been facilitated by the evolution toward domain and zonal architecture. From the viewpoint of software and programming paradigms, decoupling the functionality from the hardware has been enabled by the layered software architecture with RTOS, middleware, and service-oriented layers. In parallel, the capability of continuous improvements and evolutions has been made possible by OTA update pipelines, involving multiple steps of authentication and rollback protection.

Regarding the automation capabilities of the SDVs, the automated driving function includes several stages from detection to perception and motion planning. This tightly coupled pipeline relies on the combination of multi-modal sensors with fast on-board processors to ensure reliable operation. Moreover, AI technologies such as convolutional neural networks for perception tasks, recurrent neural networks for temporal reasoning, and reinforcement learning for adaptability are employed at each step of the automation pipeline. These techniques impose significant compute demands on the vehicle, directly affecting the power consumption and the design of the architecture.

The next step of the evolution of SDVs involves the adoption of SDN in the context of vehicular communication. SDIoV, or Software-Defined Internet of Vehicles, allows for centralized management of the communication infrastructure, enabling more efficient and flexible management compared to distributed communication protocols. In addition, edge and fog computing provide additional computing resources closer to the source (e.g., at the vehicle's or roadside unit's edge). Use cases of SDV range from critical applications such as collision avoidance and cooperative platooning to infotainment or other noncritical systems. Overall, many use cases rely on the SDV's low latency and scalability capabilities.

Five important SDV-related challenges have been also identified, namely managing a vast amount of data produced by the sensor suite, communication interfaces, and cloud services requires novel techniques, such as data compression, edge offloading, and secure processing. Furthermore, the increased attack surface that arises due to SDV technology poses serious cybersecurity concerns, related to OTA channels, V2X connections, infotainment systems, and perception systems based on deep learning models. Standardization of communication interfaces, architectures, and security protocols remains crucial for the SDV interoperability across heterogeneous platforms.

Several interesting SDV-related directions can be expected for the upcoming years. Namely, digital twins will probably become a critical part of the OTA validation and predictive maintenance process. Federated learning might be used for improving AI models collaboratively. Moreover, proactive cybersecurity measures should be taken into account, including lightweight anomaly detection models and intrusion detection systems deployed at the edge. Lastly, the advent of the large language model-based interfaces might be followed by AI-defined vehicles with autonomous decision-making abilities.

{\footnotesize
\setlength{\itemsep}{0pt}
{\footnotesize\begin{singlespace}
\setlength{\itemsep}{0pt}
\setlength{\parsep}{0pt}

\bibliographystyle{IEEEtran}
\bibliography{refs}
\end{singlespace}}
}

\vspace{-12pt}\begin{IEEEbiography}[
{\includegraphics[width=0.75in,height=0.95in,clip,keepaspectratio]{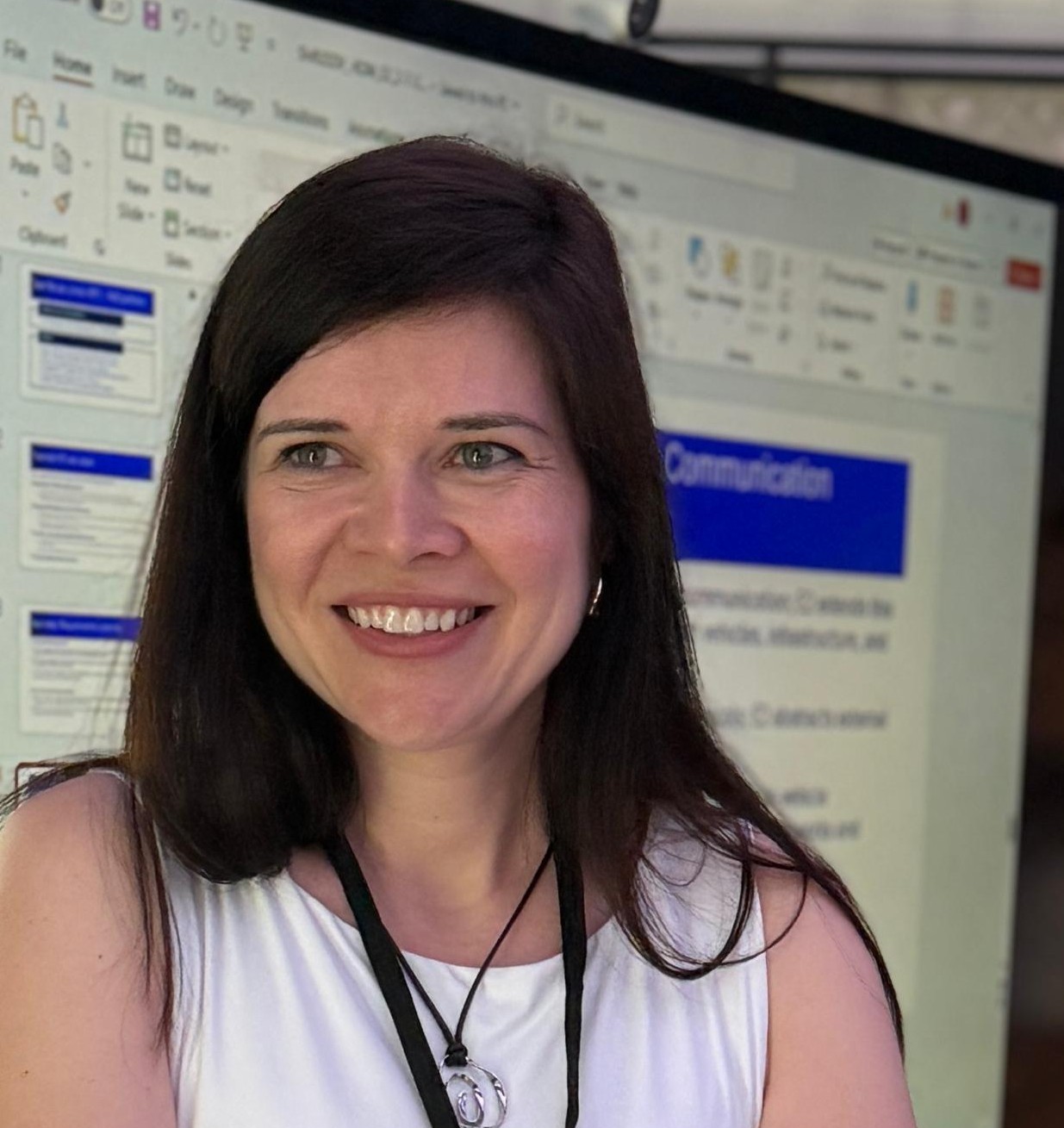}}]{Eirini Liotou }is an Assistant Professor at the Department of Informatics and Telematics, Harokopio University of Athens (since 2024). She holds a PhD from the National and Kapodistrian University of Athens, an MSc from Imperial College London (Communications and Signal Processing), an MSc in Informatics and Telecommunications, and a Diploma in Electrical and Computer Engineering from NTUA. She has worked as a Senior Software Engineer at Siemens Enterprise Communications, Post-Doc researcher, and Scientific Project Manager for EU projects at ICCS. She has authored more than 30 peer-reviewed publications and serves as an EC expert reviewer. Her research interests include 6G networks, SDN, NFV, AI in networking, SDVs, and V2X communications.
\end{IEEEbiography}

\vspace{-12pt}\begin{IEEEbiography}[
{\includegraphics[width=0.75in,height=0.95in,clip,keepaspectratio]{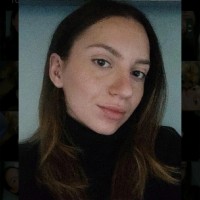}}]{Dimitra Tzelalidou }is a software engineer working at Netcompany-Intrasoft, where she contributes to the development and implementation of software solutions within large-scale information technology projects. She holds a degree from Harokopio University of Athens, where she developed a strong academic foundation in computer science and information systems. Her professional work focuses on software engineering practices, system development, and collaborative project delivery in enterprise-level digital environments, reflecting experience in modern software technologies and structured development workflows.
\end{IEEEbiography}

\vspace{-12pt}\begin{IEEEbiography}[
{\includegraphics[width=0.75in,height=0.95in,clip,keepaspectratio]{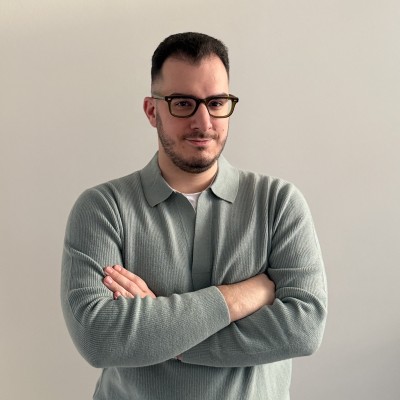}}]{Gerasimos Christodoulou }is a software engineer with professional experience in the information technology and online services sector. He has worked as a Software Engineer at Novibet, contributing to the development and maintenance of large-scale software systems within a high-performance digital environment. Christodoulou holds a degree from Harokopio University of Athens, where he developed a foundation in computing and information technologies. His professional work focuses on software development practices, system performance, and modern application architectures, reflecting experience in collaborative engineering environments and production-level software systems. 
\end{IEEEbiography}

\onecolumn
\begin{center}
\begin{longtable}{@{}ll@{}}
\caption{List of Abbreviations} \label{tab:abbreviations} \\
\toprule
\textbf{Abbreviation} & \textbf{Definition} \\
\midrule
\endfirsthead

\multicolumn{2}{c}{\tablename\ \thetable{} -- continued from previous page} \\
\toprule
\textbf{Abbreviation} & \textbf{Definition} \\
\midrule
\endhead

\bottomrule
\endfoot

\bottomrule
\endlastfoot

AAOS & Android Automotive Operating System \\
ACC & Adaptive Cruise Control \\
AEB & Autonomous Emergency Braking \\
AGL & Automotive Grade Linux \\
AI & Artificial Intelligence \\
AIDV & AI-Defined Vehicle \\
ASIL & Automotive Safety Integrity Level \\
AUTOSAR & Automotive Open System Architecture \\
BFT & Byzantine Fault Tolerant \\
BS & Base Station \\
CAN & Controller Area Network \\
CNN & Convolutional Neural Network \\
COVESA & Connected Vehicle Systems Alliance \\
C-V2X & Cellular Vehicle-to-Everything \\
DSA & Dynamic Spectrum Access \\
DSRC & Dedicated Short-Range Communication \\
E/E & Electrical/Electronic \\
EC-SDIoV & Edge Computing-Enabled Software-Defined Internet of Vehicles \\
ECU & Electronic Control Unit \\
EKF & Extended Kalman Filter \\
ESP & Electronic Stability Program \\
ETSI & European Telecommunications Standards Institute \\
FC-SDIoV & Fog Computing-Enabled Software-Defined Internet of Vehicles \\
FCW & Forward Collision Warning \\
FOTA & Firmware Over-the-Air \\
GDPR & General Data Protection Regulation \\
GIS & Geographic Information System \\
GNSS & Global Navigation Satellite System \\
GPS & Global Positioning System \\
GRU & Gated Recurrent Unit \\
HD & High-Definition \\
HDV & Hardware-Defined Vehicle \\
HMAC & Hash-based Message Authentication Code \\
HMI & Human-Machine Interface \\
HPCU & High-Performance Computing Unit \\
IDS & Intrusion Detection System \\
IMU & Inertial Measurement Unit \\
IoT & Internet of Things \\
IoV & Internet of Vehicles \\
ISO & International Organization for Standardization \\
ITS & Intelligent Transportation Systems \\
IVI & In-Vehicle Infotainment \\
LIN & Local Interconnect Network \\
LLM & Large Language Model \\
LSTM & Long Short-Term Memory \\
LTE & Long-Term Evolution \\
MaaS & Mobility as a Service \\
MEC & Multi-Access Edge Computing \\
NLP & Natural Language Processing \\
NR & New Radio \\
OBD & On-Board Diagnostics \\
OBU & On-Board Unit \\
OS & Operating System \\
OTA & Over-the-Air \\
POSIX & Portable Operating System Interface \\
QoS & Quality of Service \\
RL & Reinforcement Learning \\
RSU & Roadside Unit \\
RTE & Runtime Environment \\
RTK & Real-Time Kinematic \\
RTOS & Real-Time Operating System \\
SAE & Society of Automotive Engineers \\
S-CORE & Safe Open Vehicle Core \\
SDIoV & Software-Defined Internet of Vehicles \\
SDN & Software-Defined Networking \\
SDV & Software-Defined Vehicle \\
SLAM & Simultaneous Localization and Mapping \\
SNN & Spiking Neural Network \\
SoC & System-on-Chip \\
SOA & Service-Oriented Architecture \\
SOAFEE & Scalable Open Architecture for Embedded Edge \\
SOC & Service-Oriented Communication \\
SOME/IP & Scalable service-Oriented MiddlewarE over IP \\
SOTA & Software Over-the-Air \\
SOTIF & Safety of the Intended Functionality \\
TCU & Telematics Control Unit \\
TinyML & Tiny Machine Learning \\
UNECE & United Nations Economic Commission for Europe \\
V2I & Vehicle-to-Infrastructure \\
V2P & Vehicle-to-Pedestrian \\
V2R & Vehicle-to-Roadside \\
V2V & Vehicle-to-Vehicle \\
V2X & Vehicle-to-Everything \\
VaaS & Vehicle as a Service \\
VR & Virtual Reality \\
VSS & Vehicle Signal Specification \\
WAVE & Wireless Access in Vehicular Environments \\
WGS 84 & World Geodetic System 1984 \\

\end{longtable}
\end{center}
\twocolumn

\end{document}